\documentclass[aps, prx, twocolumn, superscriptaddress]{revtex4-2}

\usepackage[utf8]{inputenc}
\usepackage{amsmath}
\usepackage{amssymb,amsfonts,latexsym}
\usepackage{bm}
\usepackage[mathcal]{euscript}
\usepackage{graphicx}
\usepackage{epsfig}
\usepackage{color}
\usepackage{pifont,wasysym,marvosym}
\usepackage{textcomp}
\usepackage{comment}
\usepackage{epstopdf}
\usepackage{gensymb}
\usepackage{dcolumn}
\usepackage{hyperref}
\usepackage{physics}
\usepackage{dsfont}
\begin{document}
\title{Non-reciprocal breathing solitons}

\author{Jonas Veenstra}
\thanks{These authors contributed equally}
\affiliation{Institute of Physics, Universiteit van Amsterdam, Science Park 904, 1098 XH Amsterdam, The Netherlands}
\author{Oleksandr Gamayun}
\thanks{These authors contributed equally}
\affiliation{London Institute for Mathematical Sciences, Royal Institution, 21 Albemarle St, London W1S 4BS, UK}
\author{Martin Brandenbourger}
\thanks{These authors contributed equally}
\affiliation{Institute of Physics, Universiteit van Amsterdam, Science Park 904, 1098 XH Amsterdam, The Netherlands}
\affiliation{Aix-Marseille Universit\'e, CNRS, Centrale M\'editerran\'ee, IRPHE, UMR 7342, 13384 Marseille, France}
\author{Freek van Gorp}
\affiliation{Institute of Physics, Universiteit van Amsterdam, Science Park 904, 1098 XH Amsterdam, The Netherlands}
\author{Hans Terwisscha-Dekker}
\affiliation{Institute of Physics, Universiteit van Amsterdam, Science Park 904, 1098 XH Amsterdam, The Netherlands}
\author{Jean-Sébastien Caux}
\affiliation{Institute of Physics, Universiteit van Amsterdam, Science Park 904, 1098 XH Amsterdam, The Netherlands}
\author{Corentin Coulais}
\affiliation{Institute of Physics, Universiteit van Amsterdam, Science Park 904, 1098 XH Amsterdam, The Netherlands}

\begin{abstract}
Breathing solitons consist of a fast beating wave within a compact envelope of stable shape and velocity. They can propagate and carry information and energy
in a variety of contexts such as plasmas, optical fibers and cold atoms, but propagating breathers have remained elusive when energy conservation is broken.
Here, we report on the observation of breathing, unidirectional, arbitrarily long-lived solitons in non-reciprocal, non-conservative active metamaterials. Combining precision desktop experiments, numerical simulations and perturbation theory on generalizations of the sine-Gordon and nonlinear Schrödinger equations, we demonstrate that unidirectional breathers generically emerge in weakly nonlinear non-reciprocal materials, and that their dynamics are governed by an unstable fixed point. Crucially, breathing solitons can persist for arbitrarily long times provided: (i) this fixed point displays a bifurcation when a delicate balance between energy injection and dissipation is struck; (ii) the initial conditions allow the dynamics to reach this bifurcation point. Importantly, discrete effects tend to stabilize these non-reciprocal breathers over a wider range of initial conditions.
Our work establishes non-reciprocity as a promising avenue to generate stable nonlinear unidirectional waves, and could be generalized beyond metamaterials to optics, soft matter and superconducting circuits.
\end{abstract}

\maketitle

\section{Introduction}

Breaking the symmetry of interactions is commonplace in complex systems as diverse as prey-predator interactions~\cite{Lotka,volterra1926fluctuations}, networks of neurons~\cite{Neurons1986}, and human behavior~\cite{Nagatani_RepProgPhys2002,Strogatz_PRL2011}. This idea has recently seen an explosion of activity across physics. Non-reciprocity has been
theoretically introduced in synthetic many-body systems such as quantum~\cite{Hatano_PRL1996,Kunst_PRL2018,Gong_PRX2018,MartinezAlvarez_PRB2018,Yao_PRL2018,McDonald_NatComm2020} and spin systems~\cite{RyoArxiv}, and experimentally realized in optical devices~\cite{Mathew_Nat_Nanotech2020,DelPino_Nature2022,Zhong_NatPhys2020,Weidemann_Science2022}, plasmonic nanoparticles under optical drive~\cite{ZheludevNatPhys}, colloids~\cite{Bililign2021Motile,TanNature}, emulsions~\cite{Poncet_PRL2022,Colen2024-ot,Guillet2025-ap}, and metamaterials~\cite{brandenbourger2019non,Rosa_NJP2020,Scheibner2020non,Chen_NatComm2021,Wang2022,Veenstra_Nature2024,Veenstra_Nature2025}. In these systems, non-reciprocity 
is typically achieved by injecting momentum or energy in a way that breaks the symmetry or Hermiticity of the linear operators governing the dynamics~\cite{Coulais_NatPhys2021,Bergholtz_RMP2021,Shankar_NatRevPhys2022} and entails that the transfer function differs if you swap input and output.
Such asymmetry then often leads to unidirectional or chiral amplification, necessarily pushing the dynamics beyond the linear regime
into one where nonlinearities dominate and features like solitons can emerge.
Nonlinear non-reciprocal dynamics have recently been investigated in the context of pattern formation~\cite{Marchetti_PNAS2020,Saha_PRX2020,Fruchart_Nature2021}, topological excitations and defects~\cite{Bililign2021Motile,Braverman2021,TanNature,Veenstra_Nature2024}, shocks~\cite{Colen2024-ot}, turbulence~\cite{deWit_Nature_2024}, and topological edge modes~\cite{hohmann2022observation,Many_Manda2024-nb}. Yet, non-reciprocal dynamics of localized non-topological nonlinear excitations such as solitons have remained largely unexplored.

Non-topological solitons are stable and localized weakly nonlinear excitations
~\cite{dauxois2006physics} and can exist in media that are driven far from equilibrium~\cite{Book_dissipativesolitons}, in which case they are often called dissipative solitons. Common examples of such driven-dissipative media include plasmas~\cite{Kuznetsov1977}, optical microcavities~\cite{Lucas_NatComm2017,Yu_NatComm2017,Pernet_NatPhys2022}, granular materials~\cite{Nesterenko_1984, Daraio_PRE2006,Spadoni_PNAS2010}, and mechanical metamaterials~\cite{Deng_PRL2017,Deng2018,Deng2019,Deng_JMPS2021,Deng_review2021, Kochmann_Bertoldi_review, Chen2014,Nadkarni_PRE2014, Nadkarni_PRL2016, Nadkarni_PRB2016, Raney2016, Deng_PRL2017, Xiaofei_Nature2023,Janbaz_NatComm2024, Veenstra_Nature2024}. in contrast to solitons in conservative systems, dissipative solitons must strike a balance between loss and gain to be stable.
in contrast to topological dissipative solitons, non-topological solitons must strike an additional balance between dispersion and nonlinearity. These two balances put severe constraints on the lifetime and dynamics of dissipative solitons~\cite{Book_dissipativesolitons,Grelu_NatPhot2012}. A natural set of questions hence arises: What types of solitons emerge in weakly nonlinear non-reciprocal materials? Which mechanisms govern their dynamics?

Here, we address these questions by focusing on breathing solitons.
Breathing solitons are localized nonlinear excitations, modulated by a carrier wave. They can either be standing---in which case they are sometimes called oscillons---or traveling---in which case they can be used to transport information and energy.
Standing breathing solitons have been observed in a plethora of dissipative systems, from optical cavities~\cite{Parra-Rivas_PRA2014,Lucas_NatComm2017,Yu_NatComm2017} and arrays of nanoplasmonic particles in the presence of driving and dissipation~\cite{Noskov_PRL2012,Noskov_SciRep2012,Noskov_RSPA2014,Hu_PRR2024}, to driven granular media~\cite{Umbanhowar_Nature1996} and fluids~\cite{Fineberg_PRL}. In stark contrast, traveling breathing solitons have yet to be observed in strongly dissipative systems. The only exception can be found in nanoplasmonic particle arrays, where traveling patterns reminiscent of breathing solitons have been observed numerically under a unidirectional external drive~\cite{Noskov_SciRep2012}, yet without experiments or theoretical description of the solitons.

\begin{figure}[t!]
\centering
\hspace{0in}
\includegraphics[width=1.0\columnwidth,trim=0cm 0cm 0cm 0cm]{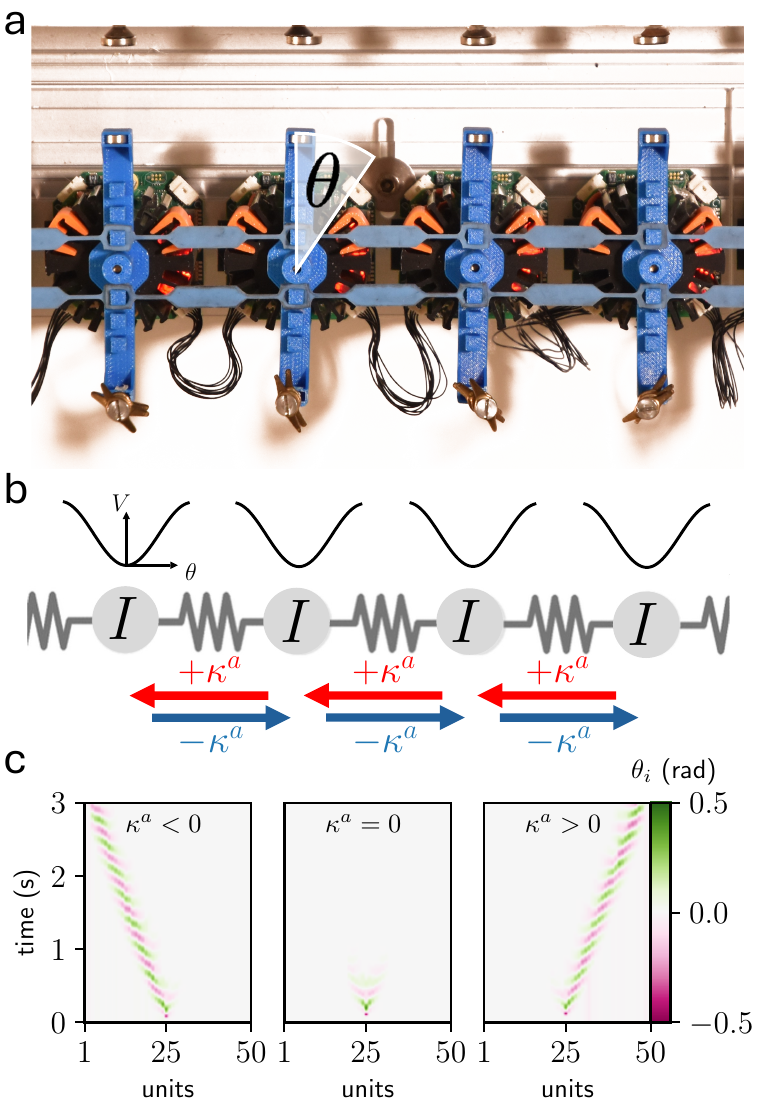}
\caption{
\textbf{Breathing non-reciprocal soliton.} 
{\bf (a)} Picture of four identical unit cells composing the active mechanical metamaterial. Each unit has one rotational degree of freedom (vertical blue bar) which is elastically coupled to its nearest neighbors via the horizontal thin blue elastic bands. The non-reciprocal gain is programmed by the microcontroller embedded in each unit (Appendix \ref{Experimental Methods}). Magnets placed on the oscillator tips and a periodic substrate generate a weakly nonlinear potential. The moment of inertia is controlled by weights to the oscillator tips (golden bolts). 
{\bf (b)} Schematic overview of the metamaterial whose rotational degrees of freedom have moment of inertia $I$, are connected by springs offset by non-reciprocity $\kappa^a$ (red/blue arrows) and are each subject to a weakly nonlinear magnetic potential $V$.
{\bf (c)} 
Kymograph of the angular deviations $\theta_i$ of 50 connected oscillators upon excitation of the middle oscillator. Without non-reciprocal gain $\kappa^a$, the excitation quickly disperses and dies out (middle panel).
For positive (negative) gain, a breathing localized wave forms and travels to the right (left),
as seen in the right (left) panel.
}
\label{Fig1}
\end{figure}

We discover here that traveling breathing solitons can spontaneously emerge in non-reciprocal media. 
Using experiments on active metamaterials, numerical simulations and perturbation theory, 
we find that a delicate balance between initial conditions, non-reciprocal gain and damping sets the conditions for the emergence and lifetime of these solitons. This balance determines whether radiation, one breather, multiple breathers or kink-antikink pairs are formed following an initial perturbation; together with the magnitude and velocity of the initial condition, this balance also determines fate of the breathing soliton.

Strikingly, we find that in contrast to breather solitons in the presence of constant driving~\cite{Parra-Rivas_PRA2014,Lucas_NatComm2017,Yu_NatComm2017} (see Appendix~\eqref{NLSconstant}), non-reciprocity makes the dissipative breathers travel and expands the regime of existence of traditional breathing dissipative solitons. Our results establish active metamaterials as a productive avenue for the study of unidirectional dissipative solitons and for nonlinear generalizations of non-reciprocal excitations.

\section{Breather-like wave in a non-reciprocal active metamaterial}

Consider the active mechanical metamaterial shown in Fig.~\ref{Fig1}a and in Supplementary Movie 1~\cite{SuppVideo1}. It is made of 50 oscillators, each linked to its neighbors by a thin flexible rubber plate (see Appendix \ref{Experimental Methods} for details). Upon relative rotation of neighboring oscillators $\theta_i$ and $\theta_{i+1}$, this plate will bend and stretch and thereby apply a nonlinear elastic restoring torque on oscillator $i$, $\tau^e_i(\theta_i,\theta_{i+1})$. This restoring torque is symmetrical by virtue of Newton's third law $\tau^e_i(\theta_i,\theta_{i+1})=- \tau^e_{i+1}(\theta_i,\theta_{i+1})$. By design, we break Newton's third law by applying an external torque on each oscillator. This torque depends on the configuration of the system $\tau^a_i=\kappa^a(\theta_{i-1}-\theta_{i+1})$, where increasing $i$ corresponds to the clockwise direction. This is done by using an active control system that relies on motors, sensors and microcontrollers embedded in our oscillators. This active torque introduces a non-reciprocal bias in the system and amplifies waves and pulses unidirectionally, in what is known as the non-Hermitian skin effect~\cite{brandenbourger2019non,Coulais_NatPhys2021,Bergholtz_RMP2021,Shankar_NatRevPhys2022}. As unidirectional amplification occurs, the magnitude of the signal may increase and nonlinearities will inevitably come into play. 
In order to probe the nonlinear response in a controlled way, we place magnets both on the oscillator tips and periodically on a substrate \cite{Veenstra_Nature2024}, which creates a weakly nonlinear potential (Fig.\ref{Fig1}b and see Appendix~\ref{Experimental Methods}). 
We first probe the response of the active mechanical metamaterial to an initial perturbation which we apply manually (see Supplementary Video 1).
Without non-reciprocity $\kappa^a=0$,  the perturbation disperses symmetrically around the locus of impact and dissipates within a fraction of a second (Fig.\ref{Fig1}c). For nonzero $\kappa^a$, we observe that the perturbation transforms into a localized and oscillating wave packet that propagates at a constant velocity of $8 \pm 1$ units/s in a direction given by the sign of $\kappa^a$. Strikingly, non-reciprocity mitigates the dispersive and dissipative effects and gives rise to a traveling and localized oscillating wave whose lifetime can be extended to multiple seconds depending sensitively on the strength of $\kappa^a$. Besides the stable envelope, this wave has a carrier wave that is oscillating at a frequency of $5$ Hz and moving at a constant phase velocity of $28$ units/s.

\begin{figure}[t!]
\centering
\includegraphics[width=1.0\columnwidth,trim=0cm 0cm 0cm 0cm]{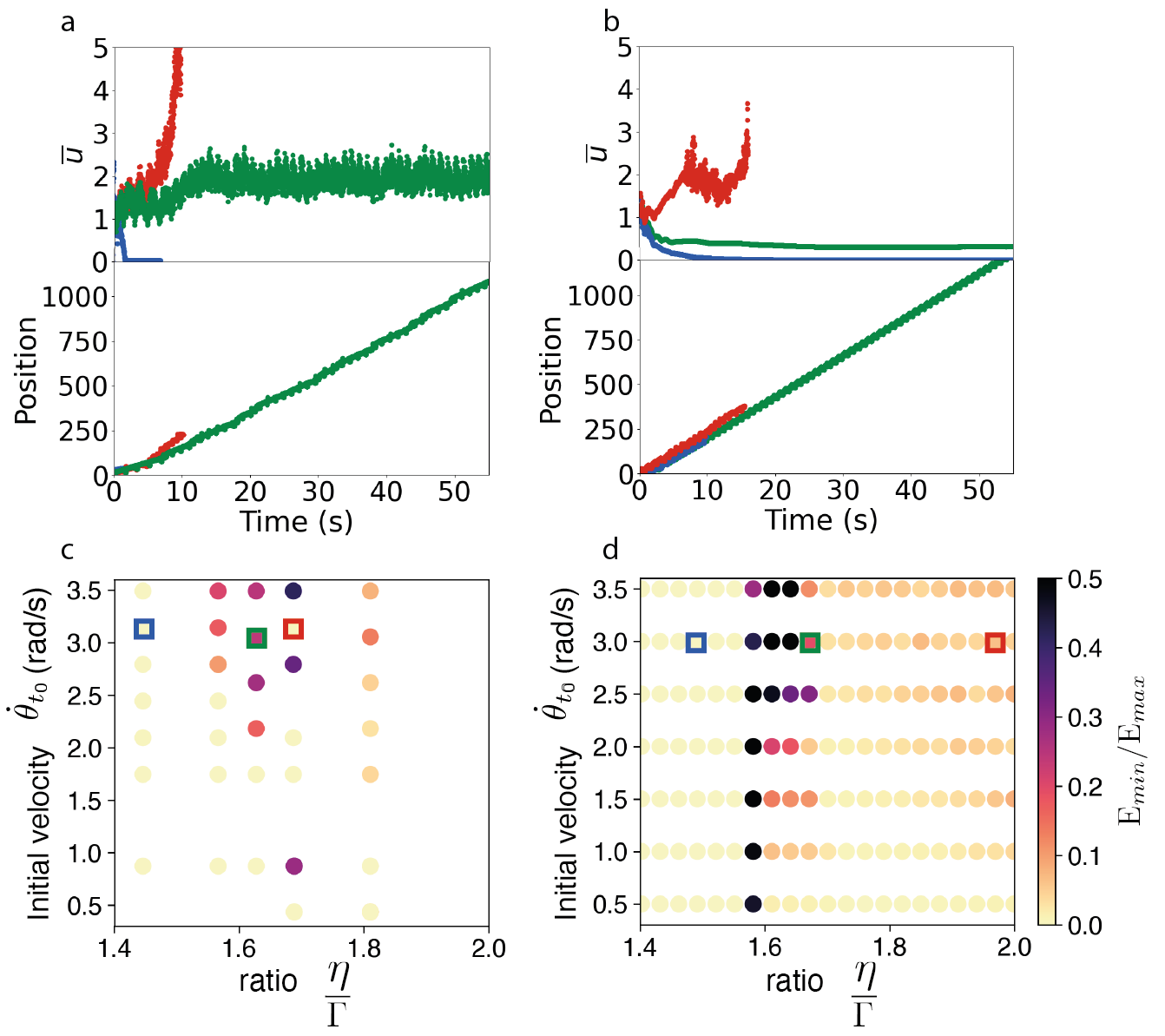}
\caption{\textbf{Three regimes of breathers.}
{\bf (a)} The total squared amplitude of the oscillators over time for a range of the non-dimensional gain $\eta$ normalized by the viscous dissipation $\Gamma = 0.3 \pm 0.1$ shows decaying (blue), diverging (red) and long lived (green) breathers. Black line shows the passive case $\eta=0$. As a metric for the longevity of the breather, we use the ratio $E_{\mathrm{min}}/E_{\mathrm{max}}$ between minimum and maximum energy measured within a window demarcated by the dotted lines.
{\bf (b)} A sweep over initial conditions $\dot{\theta}_{t=0}$ and $\eta/\Gamma$ shows a narrow regime of extended breather lifetime, quantified by the logarithm of the ratio between the minimum and maximum amplitudes of the breather measured within a time period of one second, indicated by the dashed vertical lines in panel {\bf (a)} (see Methods Section \ref{Experimental Methods}). Square and circularly hatched areas demarcate regions of stable decaying and unstable amplifying waves respectively.
{\bf (c)} Simulations of Eq.~\eqref{eq:NRFK} using experimental values $\Gamma=0.3$ and $D=1.9$.
{\bf (d)} A sweep over the same initial conditions reveals a narrow range of breathers with an extended lifetime. 
}
\label{Fig2}
\end{figure}

Under which conditions can nonreciprocity sustain this nonlinear excitation? 
To ascertain this, we perform a set of experiments in which we systematically vary the non-dimensional gain $\eta \propto \kappa^a/\kappa$ and the perturbation velocity $\dot{\theta}_{t=0}$ (Fig.~\ref{Fig2}ab), now controlled by a pulse exerted by the units internal motor. 
This reveals the strong influence of the non-reciprocal gain:
when the gain is too small, the perturbation disperses asymmetrically but quickly dissipates; when $\eta$ is too large, oscillators are rendered unstable and a rapidly amplifying wave emerges.

Only when the non-reciprocity is carefully tuned,
can a localized wave of constant amplitude be observed that lasts several seconds. 

Why do these waves emerge from non-reciprocal interactions? And what is their nature?
To answer these questions, we first consolidate our experimental findings using a discrete model that captures all the essential features of the non-reciprocal active material. We model the dynamics of the interacting oscillators subject to a nonlinear potential by a non-reciprocal version of the Frenkel-Kontorova model, which reads in non-dimensional form (see Appendix \ref{Numerical Methods}):
    \begin{align}
    \Ddot{\phi}_i\!= \! \phi_{i-1}\!+\!\phi_{i+1}\!-\!2\phi_{i}\!-\!\frac{\eta}{2} (\phi_{i+1}\! -\! \phi_{i-1}) \! -\! \Gamma \dot{\phi_i} \!-\! D \sin(\phi_i) 
    \label{eq:NRFK}
    \end{align}
where, $\phi_i=2\pi\frac{\theta_i}{\theta_d}$ denotes the $i^\mathrm{th}$ oscillator angle normalized by the period of the magnetic potential $\theta_d=1 \,\mathrm{rad}$. The nondimensional parameters $\Gamma=0.3\pm0.1$ and $D=1.9\pm0.3$ represent on-site dissipation of the oscillators and magnetic potential strength and $\eta \propto \kappa^a/\kappa$ whose values we calibrate experimentally using a torsion testing machine (see Methods Section \ref{Experimental Methods} for details).
We numerically integrate this equation and find a quantitative agreement with the experiments (Fig.~\ref{Fig2}b-d)
with additional insights from the greater precision of our numerical simulations. 
First, the velocity of the breathing waves does not sensitively depend on the level of non-reciprocity (Fig.~\ref{Fig2}b). 
Second, long-lasting breathing waves exist only for a narrow regime of non-reciprocity and in this regime they emerge for a wide range of perturbation velocities. 
Higher velocity perturbations extend the region of stable solutions to a wider range of the non-reciprocal gain.
This suggests that the existence of long-lasting unidirectional nonlinear waves depends on a precise balance between gain and loss.

\section{Non-reciprocal breathing solitons}
\label{sec:sineG}

To confirm that a nonlinear non-reciprocal material can host breathing solitons, we take the continuum limit of Eq.~\eqref{eq:NRFK} and find a sine-Gordon equation (see Methods Section \ref{Numerical Methods}), augmented with non-reciprocity and damping:
    \begin{equation}
    \varphi_{tt} - \varphi_{xx} + \text{sin}(\varphi) =  -\eta \varphi_x - \Gamma \varphi_t \label{sineG}.
\end{equation}
While a related version of this equation with a constant driving instead of a non-reciprocal term has been studied in detail~\cite{McLaughlinScott1978,dauxois2006physics}, breathers in its non-reciprocal counterpart have to the best of our knowledge remained unexplored. Crucially, the left-hand side of Eq.~\eqref{sineG} is integrable, which will allow us to
unambiguously identify the breathers (as well as the other discrete structures) and explore their full evolution in Eq.~\eqref{sineG} by means of the perturbation theory applied to the auxiliary linear scattering problem.

\begin{figure*}[t!]
%\hspace{-2.2cm}
\centering
\hspace{0in}
\includegraphics[width=2.\columnwidth,trim=0cm 0cm 0cm 0cm]{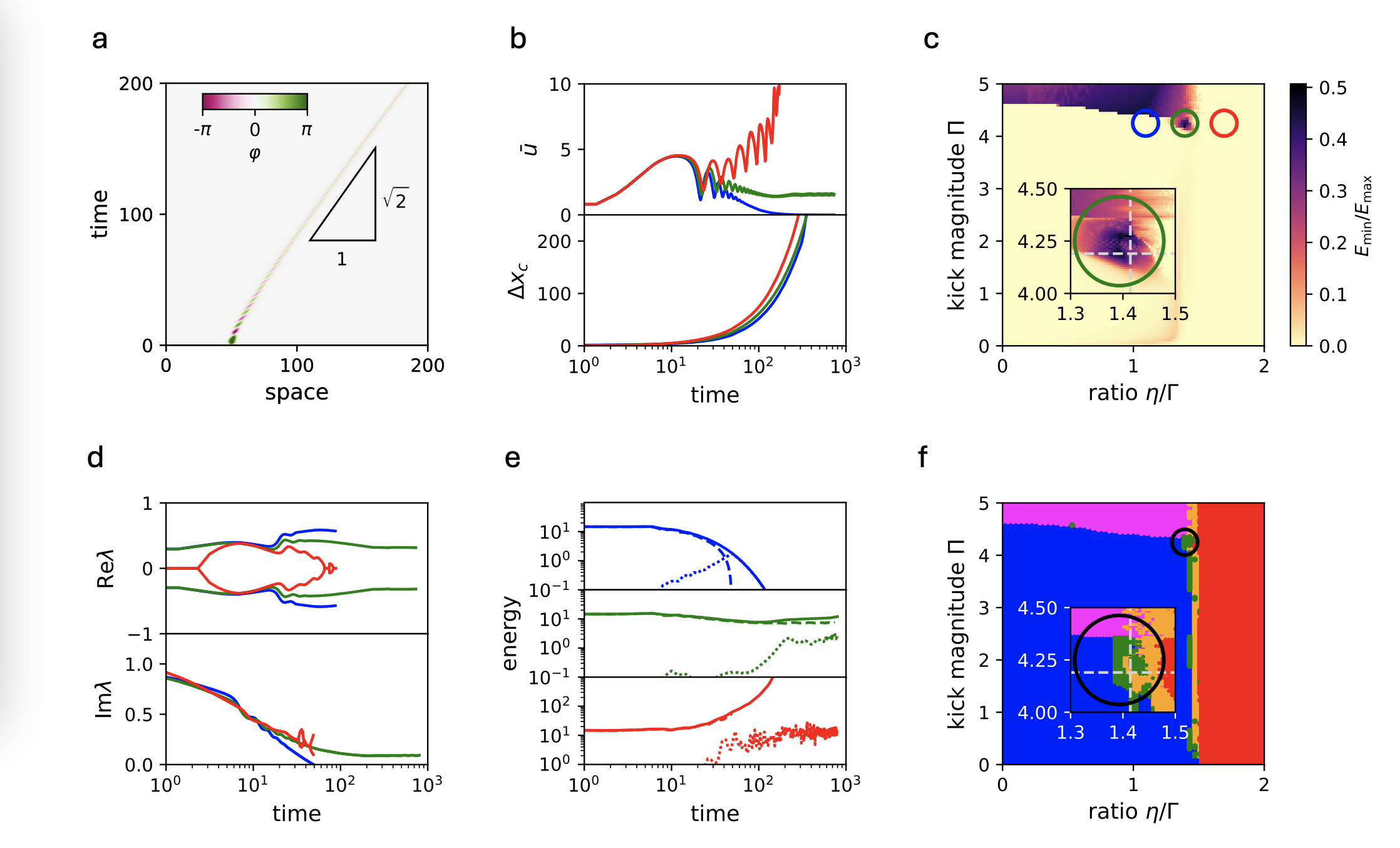}
\caption{\textbf{Breathers in the non-reciprocal sine-Gordon equation.}
{\bf (a)} Kymograph of the field $\varphi$ for $\eta=0.1373$, $\Gamma=0.1$ and $\Pi =4.17$ through a numerical integration of Eq.~(\ref{sineG}).
{\bf (b)} Total energy $\mathcal{E}$ of the breather (top) as defined by Eq.~\eqref{EEE} and the position of breather $\Delta x_{c}$ versus time (bottom) for three different values of non-reciprocity, $\eta = 0.10$ (blue), $0.1373$ (green) and $0.15$ (red).
{\bf (c)} Logarithm of the ratio between the minimum and maximum energy measured in the time interval indicated by the vertical dashed lines in panel (b) against the non-reciprocity to dissipation ratio $\eta/\Gamma$ and the magnitude of the initial kick $\Pi$. The energy is defined in Eq.~\eqref{EEE}.
The three colored circles correspond to the three different values plotted in panel (b). Regions hatched with squares (circles) demarcate the weakly (strongly) nonlinear regime of Eq.~\eqref{sineG}.(inset) close up on the region that hosts long-lived breathers. 
{\bf (d)} Time dependence of the real (top) and imaginary (bottom) parts of $\lambda_1$ and $\lambda_2$ for the same values of $\Pi$ and $\eta$.
{\bf (e)} Energy dependence of the soliton profile for the same values of $\Pi$ and $\eta$. Solid lines represent the energy of the field found by numeric integration ~\eqref{EEE}, and dashed (dotted) lines are contributions from the discrete (radiative) modes only, as given by Eq.~\eqref{d2}.
{\bf (f)} Spectral content of the nonlinear wave vs. ratio between non-reciprocity and dissipation $\eta/\Gamma$ and magnitude of the initial kick $\Pi$ at time $t=800$. The narrow green area shows the region of the parameters which correspond to a breather and radiation modes. This area separates decaying profiles consisting solely of radiation modes (blue region) and the regions marked in red and orange that correspond to the diverging profiles and one or more discrete degrees of freedom respectively. The orange region corresponds to a situation when additionally to a breather other discrete degrees of freedom are present in the profile (i.e. more breathers or, even, kink-antikink pairs).
The full energy in this case still should be low enough to compute $a(\lambda)$ and extract the corresponding zeroes (contrary to the red region). Notice that the ``upper'' boundary separates breathers from the kink-antikink pairs (orange area). Persistent breathers only live in a tiny region (see inset) of the parameter space $\eta/\Gamma\simeq 1.4$ and $\Pi\simeq 4.2$ indicated by the white circle.}
\label{Fig3}
\end{figure*}

We perform a numerical investigation of profile dynamics governed by Eq.~\eqref{sineG} for a fixed value of the damping $\Gamma=0.1$ and various
parameters of the non-reciprocity $\eta$. We use periodic boundary conditions, in accordance with the experiments and numerical simulations above. To simulate the ``kick'' initial condition, we impose
\begin{equation}\label{sGprofile}
	\varphi(x,0) = 0 , \qquad \partial_t \varphi(x,0) = \frac{\Pi}{\cosh(x/d)}.
\end{equation}
We have also tried other forms of kicks including a step function and a Gaussian profile and found that %qualitatively all results
breathers consistently emerge provided a balance between non-reciprocity and damping exists, and provided a sufficiently large kick is given.
%are the same provided that the injected energy is the same.
For our chosen profile Eq.~\eqref{sGprofile}, the injected energy is controlled by the magnitude $\Pi$ of the applied kick and its width $d$ which is set to $1$ in all simulations.

Similarly to the discrete case and to the experiment, we find qualitatively the same breather-like behavior (Fig.~\ref{Fig3}a) with steady a envelope and fast oscillating carrier wave. We also observe the three regimes identified earlier, where for a relatively small non-reciprocity to dissipation ratio $\eta/\Gamma=1.00$, the wave decays; when this ratio becomes larger $\eta/\Gamma=1.50$, the wave blows up and in between at $\eta/\Gamma=1.37$ the wave survives for a remarkably long time (Fig.~\ref{Fig3}b). In agreement with experiment, we find that the velocity of the wave does not depend on non-reciprocity (Fig.~\ref{Fig3}b), contrary to the case of topological solitons~\cite{Veenstra_Nature2024}.

A systematic investigation of the instantaneous energy of the profile
 \begin{equation}\label{EEE}
     \mathcal{E} = \frac{1}{2}\int \left(
     \frac{(\partial_t\varphi)^2+(\partial_x\varphi)^2}{2} + 1 - \cos\varphi
     \right) dx
 \end{equation}
at every moment of time allows us to compute the maximum variation in energy of the nonlinear wave over the course of time, and reveals that there are two regions where the nonlinear wave is particularly long-lived: (i) a specific region $\eta/\Gamma\simeq 1.4$ and $\Pi\simeq 4.2$ (Fig.~\ref{Fig3}c-inset) and (ii) a large region $\Pi>4.4$ and $\eta<1.4$.

But are these nonlinear waves breathing solitons?
To address this question, we perform a spectral decomposition according to the left-hand side of Eq.~(\ref{sineG}). Indeed, the unperturbed part is an integrable equation that can be solved exactly by the technique based on the inverse scattering transform (IST)\cite{RevModPhys.61.763,Faddeev_1987}.  This procedure can be understood as a nonlinear Fourier transformation, that transforms the profile $\varphi(x)$ into the scattering data of the auxiliary linear problem with the potential that depends on $\varphi(x)$ (so-called direct scattering).
It is convenient instead of the transmission and reflection coefficients to present the data in the functions $a(\lambda)$ and $b(\lambda)$ of the spectral parameter $\lambda$. The time evolution of these variables is extremely simple (see Appendix \eqref{theor1} for details).
 
The breather appears whenever $a(\lambda)$ has zeroes in the upper half-plane (we label these points $\lambda_{1} =\zeta+i\nu $, $\nu>0$ and $\lambda_2 = - \bar\lambda_1$). The corresponding profile reads
\begin{equation}\label{phi0}
	\varphi(x,t) = -4\arctan \frac{\lambda_1-\bar{\lambda}_1}{\lambda_1+\bar{\lambda}_1} \frac{\mathfrak{g}(x,t)-\bar{\mathfrak{g}}(x,t)}{1+|\mathfrak{g}(x,t)|^2}
\end{equation}
with
\begin{equation}\label{gamma1}
	\mathfrak{g}(x,t) = \exp\left(\frac{i x(\lambda_1^2-1)}{2\lambda_1}+\frac{i t(\lambda_1^2+1)}{2\lambda_1}\right)\mathfrak{g}(0).
\end{equation}
If the right-hand side of Eq.~\eqref{sineG} were zero, the zeroes $\lambda_1$ and $\lambda_2$ would be constant complex numbers. In the present case, in the adiabatic approximation \cite{RevModPhys.61.763}, the right-hand side leads to modification of the parameters of the breather, \emph{i.e.} $\lambda_1$ and $\lambda_2$ start depending on time. This means that the features of the solitons such as amplitude and velocity evolve over time as a consequence of dissipation and non-reciprocity.

The scattering data of the initial profile Eq.~\eqref{sGprofile} can be found explicitly  (see Eqs. \eqref{scat1} and \eqref{scat2} of the Appendix). The breather is present in the initial profile if $2 < \Pi d <4$ and the corresponding zeroes are given by
\begin{equation}\label{l12}
    	\lambda_{1,2} = i\frac{\Pi d-2}{2d} \pm\sqrt{1-\left(\frac{\Pi d-2}{2d}\right)^2}.
\end{equation}
In the unperturbed system, these spectral data would correspond to the profile
\begin{equation}\label{approx22}
 	\varphi =  4 \arctan  \frac{\frac{\Pi d-2}{2d} }{\sqrt{1-\left(\frac{\Pi d-2}{2d}\right)^2}} \frac{\sin  t\sqrt{1-\left(\frac{\Pi d-2}{2d}\right)^2}}{\cosh \frac{\Pi d-2}{2d}x}.
 \end{equation}
 For the time-evolved profile the corresponding scattering data can be computed numerically. We employ the transfer-matrix method described in
 \cite{BOFFETTA1992252}.
 Strictly speaking the initial condition for the profile corresponds to a kink-antikink pair---where $\lambda_1$ and $\lambda_2$ are purely imaginary, rather than a breather---where $i\lambda_1$ and $i\lambda_2$ are conjugate complex numbers. In the simulation however, this pair immediately transforms into a breather because of the dissipative terms, which is reflected in the non-zero real part in Fig.~\ref{Fig3}d. Notice also that for large non-reciprocity the kink-antikink pair profile  might persist for short times (red curve until $t=3$) and  return at later times (red curve at time $t\simeq 80$).

 The three regimes in the scattering data are reflected as follows: the dissipative regime is marked by a vanishing imaginary part, while the real part stays finite (blue curve); the exploding regime is marked by the vanishing of both real and imaginary parts (red curve), while in the stable regimes both spectral data evolve slowly over the simulation window (green curve). This confirms the existence of a breathing soliton in the minute region $\eta/\gamma\simeq 1.4$ and $\Pi\simeq 4.2$ (see Fig.~\ref{Fig3}c-inset).

To quantify this behavior even further, we have used the scattering data $a(\lambda)$ to numerically quantify the fraction of energy of discrete structures such as solitons and breathers in each profile (see Appendix \eqref{cq})
\begin{equation}
	\mathcal{E}_d = 2\sum {\rm Im}\left( \lambda_i- \lambda^{-1}_i\right),\label{discrete_energy}
\end{equation}
which in the case of a single breather becomes
  \begin{equation}\label{d2}
    \mathcal{E}_b = 2 \,{\rm Im} (\lambda_1+\lambda_2 -1/\lambda_1-1/\lambda_2).
\end{equation}
Note that generically, more than one breather may appear, together with kinks and antikinks. We can use Eq.~\eqref{discrete_energy} to quantify the total energy of such discrete nonlinear structures. In addition, we can quantify the energy of radiation as (see Appendix \eqref{cq})
 \begin{equation}\label{erad_MT}
   \mathcal{E}_r =\int\limits_0^\infty\log\frac{1}{|a(\lambda)|^{2}} \frac{1}{\pi}d \left(\lambda - \frac{1}{\lambda}\right).
 \end{equation}
Plotting $\mathcal{E}$, $\mathcal{E}_d$ and $\mathcal{E}_r$ vs. time for a few different values of parameters (Fig.~\ref{Fig3}e) sheds new light on the three possible scenarios: (i) in the decaying regime (blue curves), the discrete degrees of freedom initially dominate $\mathcal{E}_t \approx \mathcal{E}_d\gg\mathcal{E}_r$, but vanish and are replaced by radiation for long times $\mathcal{E}_t \approx \mathcal{E}_r\gg\mathcal{E}_d$;
(ii) in the exploding regime (red curves) we see that the discrete degrees of freedom take the most part of the profile's energy $\mathcal{E}_t \approx \mathcal{E}_d \gg \mathcal{E}_r$; (iii) finally in the stable regime (green curves) we see that the discrete degrees of freedom dominate the dynamics over a long transient $\mathcal{E}_t \approx \mathcal{E}_d \gg \mathcal{E}_r$. %but that the radiation progressively starts to build up.
In other regimes for smaller magnitude of initial conditions, the radiation itself spawns discrete structures (see Appendix \eqref{cq} for a complete discussion).

By making a snapshot of the spectral content for large times $t=800$, we obtain a good picture of the nature of the nonlinear waves and identify the regime where breathing solitons persist (Fig.~\ref{Fig3}f). To obtain the color plot it was enough to deduce the number of zeroes of $a(\lambda)$ without having to determine their precise positions. The number of zeroes $N_C$ inside a domain bounded by a contour $C$ is given by the following contour integral
\begin{equation}
    N_C = \frac{1}{2\pi i }\oint_C d\lambda \frac{\partial_\lambda a(\lambda)}{ a(\lambda)},
\end{equation}
providing an efficient way to compute $N_C$ numerically. To compute the number of breathers we have chosen the contour $C$ in the complex plane $\lambda$ to be a square with the opposite vertices given by $\lambda=0.1+0.1i$ and $\lambda = 1+0.7 i$.
Those values were chosen empirically to encircle most of the zeroes.
There is a natural limitation for the spectral analysis as for the very large or very small values of $\lambda$ the computation of $a(\lambda)$ suffers from numerical instabilities.
This can be seen in particular in the energy of the corresponding profiles  \eqref{discrete_energy}. Separately we have looked for the zeroes on the imaginary line.
In terms of the zeroes, the colors mean the following: in the blue region there are no zeroes, in green there is one with the non-zero real part, for the magenta all zeroes have zero real part, in the orange region there are more than two zeroes, and the red regions have the energy too high to perform the analysis. For the green and blue areas, in order to ensure that we have found all the zeroes, we compared the total energies computed from the profile and with Eqs. \eqref{discrete_energy}  and \eqref{erad_MT}.

For low non-reciprocity to dissipation ratio $\eta/\Gamma<1.4$, we see that at long times, radiation dominates (blue region), except for large value of the kick magnitude $\Pi>4.4$, where a pair of kink and antikink dominates (magenta region). This regime can not be achieved in our experiments since the oscillators are monostable.
For large non-reciprocity to dissipation ratio $\eta/\Gamma>1.5$, at long times, multiple breathers and/or kinks dominate (red region). Only for intermediate non-reciprocity $\eta/\Gamma\simeq 1.4$ will a single breather dominate the dynamics (green region and Fig.~\ref{Fig3}f-inset).

 We have seen that breathing solitons emerge for a broad range of initial conditions but only persist for a thin range of non-reciprocity to dissipation ratio $\eta/\Gamma$. What controls this range of persistence and why is it so thin? To address these questions we will next employ a multiple-scale expansion to obtain analytical predictions for the dynamics of the envelope of such non-reciprocal breathing solitons.

\section{Envelope of the non-reciprocal breathing soliton}

The adiabatic evolution of the spectral data for the perturbed sine-Gordon equation \eqref{sineG} can, in principle, be analysed via the inverse scattering perturbation theory \cite{Kosevich_Kivshar_paperInRussian,Karpman2,RevModPhys.61.763,Veenstra_Nature2024}.
However, this analysis involves a system of four integro-differential equations and at first glance is hardly simpler than the original equation.
Therefore, prompted by our observations of Fig. ~\ref{Fig1} and Fig.~\ref{Fig3} we extract qualitative predictions by turning to a simpler model of a small-amplitude breather of the form
\begin{equation}\label{approxNLS}
    \varphi(x,t) = 2\bar{\psi}(x,t) e^{2it}  + {\rm c.c.},
\end{equation}
where $\psi(x,t)$ is a slowly-varying envelope $|\partial_t \psi| \ll |\psi|$, and for convenience we have rescaled time $t \to t/2$.
After this rescaling Eq. \eqref{sineG} takes a canonical form \footnote{In accordance with notations of \cite{Faddeev_1987}.} of non-reciprocal dissipative Nonlinear Schr\"odinger equation (NLS)
\begin{equation}\label{nls0}
	i \partial_t \psi + \psi_{xx} + 2|\psi|^2\psi =  \eta\psi_x  - i \Gamma\psi + \frac{\Gamma}{2} \partial_t\psi.
\end{equation}
This equation assumes that cubic nonlinearities are sufficient to capture the nonlinear dynamics. Our calibrations done in Appendix~\ref{Experimental Methods} (Fig.~\ref{Fig:calibration}b) prove that indeed cubic nonlinearities qualitatively capture range of deformations accessible experimentally.
In the absence of perturbations on the right-hand side, Eq.~\eqref{nls0} is integrable and the soliton $\psi_s$ describes the envelope of the breather in the laboratory frame
\begin{equation}\label{profS}
    \psi_s(x,t) = \frac{u e^{iv (x-x_c(t)) + i \Phi(t)}}{\cosh(u(x-x_c(t)))},
\end{equation}
with phase $\Phi(t)$ and center mass coordinate $x_c(t)$. In integrable dynamics these variables are linear functions of time and are named angle variables, while $u(t)$ and $v(t)$ are constants (the action variables) representing the amplitude and velocity of the envelope.
In contrast, in the non-integrable case non-reciprocity and dissipation will result in more complicated time dependence for these parameters.% of Eq.~\eqref{profS}.

We now predict how the dynamics of this soliton---the envelope of the breather---depends on non-reciprocity, dissipation and initial conditions. To this end, we note that the scattering data is now given by a zero of $a(\lambda)$ in the upper half plane $\lambda_0= v+ i u$, where the parameters $v$ and $u$ are the velocity and amplitude of the profile Eq.~\eqref{profS}.
The construction of the transfer matrix is similar to the sine-Gordon case however here we do not have any restrictions on the zeroes ({\it cf} Appendices \eqref{theor1} and \eqref{nseP}).
The time evolution of these parameters as well as the parameters $\Phi$ and $x_c$ can be obtained with the inverse scattering perturbation theory, which in the adiabatic approximation leads to (see Appendix ~\eqref{nseP})
\begin{equation}\label{ffu}
\frac{dv}{dt} = \frac{2u^2}{3} (\eta - \Gamma v)
\end{equation}
\begin{equation}\label{ffv}
	\frac{du}{dt} = 2u(\eta v-\Gamma)+ u\Gamma(u^2-v^2),
\end{equation}
and for the angle variables
\begin{equation}\label{ff1}
	\frac{dx_c}{dt} = 2v, \qquad
	\frac{d\Phi}{dt} = u^2 + v^2.
\end{equation}

\begin{figure*}[t!]
\centering
\hspace{0in}
\includegraphics[width=2\columnwidth]{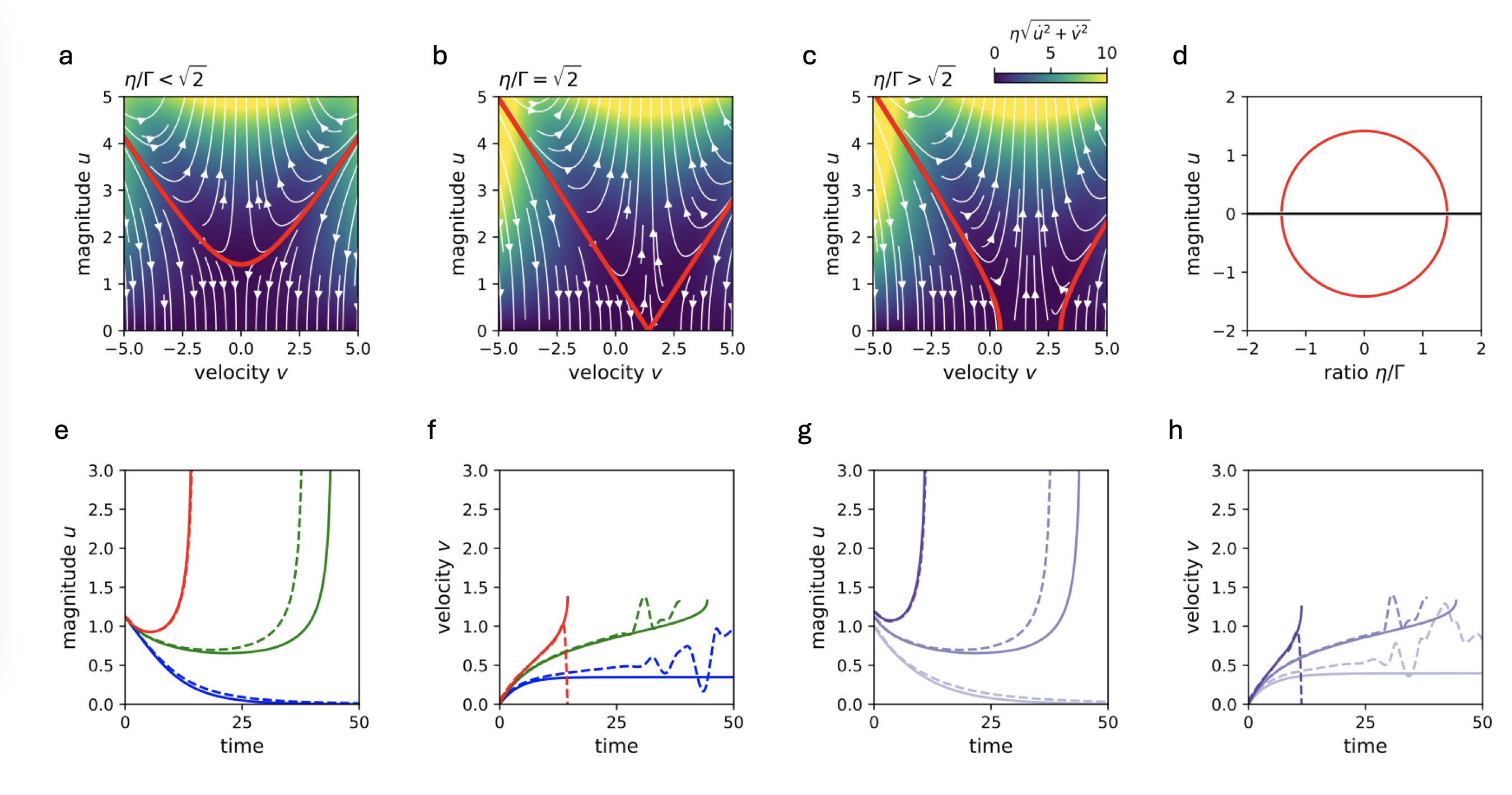}
\caption{\textbf{Long time dynamics of the envelope of the soliton.}
{\bf (a)} Phase portrait $u$ vs. $v$ for $\eta/\Gamma=0$. White streamlines and colormap indicate the direction and rate of flow respectively. The orange curve shows the separatrix given by Eq.~\eqref{separatrix} and the lightgreen cross indicates the fixed point given by Eq.~\eqref{eq:P}. {\bf (b)} Phase portrait $u$ vs. $v$ for $\eta/\Gamma=\sqrt{2}$; and
{\bf (c)} phase portrait $u$ vs. $v$ for $\eta/\Gamma=\sqrt{3}$ from Eqs.~(\ref{ffu}-\ref{ffv}). 
\textbf{{\bf (d)}} Bifurcation diagram $u$ vs. $\eta/\Gamma$. The lightgreen line indicates the pair of fixed points $P$ and the black line corresponds to the fixed point at $u=0$.
{\bf (e-f)} Trajectories $u$ (e) and $v$ (f) vs. time for $\eta=0.15$ (red), $0.1373$ (green) and $0.1$ (blue) and $\Gamma=0.1$.
Solid lines indicate numerical solutions of Eqs.~(\ref{ffu}-\ref{ffv}) and dashed lines numerical solutions of Eq.~\eqref{nls0} with Eq.~\eqref{profS} as an initial condition with $\Phi(0)=0$, and $u=1.12$.
{\bf (g-h)} Trajectories $u$ (g) and $v$ (h) vs. time for $u=1.19$ (dark purple), $1.12$ (purple) and $1.0$ (light purple) and $\Gamma=0.1$ and $\eta=0.1373$.}
\label{Fig4}
\end{figure*}

Although they are impossible to solve analytically, Eqs.~(\ref{ffu}-\ref{ffv}) reveal a physically appealing picture. First, there are two fixed points $u=0$, that corresponds to no solitons at all and $u=+\infty$ and $v=\eta/\Gamma$, which correspond to a diverging soliton. Second, in the regime $|\eta|/\Gamma\leq\sqrt{2}$ there is another pair of fixed points $P$ that corresponds to a soliton of constant velocity with a constant amplitude
\begin{equation}\label{eq:P}
v_{P}=\frac{\eta}{\Gamma} \qquad u_{P}=\pm\sqrt{2 - \frac{\eta^2}{\Gamma^2}}.
\end{equation}
To determine which fixed point governs the dynamics, we show the phase portrait in Fig.~\ref{Fig4}a-c see that the fixed point at $u=0$ is stable and attracts all the trajectories if the initial condition sits below the separatrix $S$ defined by
\begin{equation}\label{separatrix}
u_S^2 = 2 - \eta^2/\Gamma^2  + \frac{3}{5}\left(v+ \eta/\Gamma\right)^2
\end{equation}
and shown as orange curves.
In the case $|\eta|/\Gamma\leq\sqrt{2}$, if the initial conditions sit above the separatrix $S$, the trajectory starts to be attracted towards the fixed point $P$ (Fig.~\ref{Fig4}ab). The envelope of the breather shrinks in magnitude towards a magnitude $u_{P}$ and its velocity converges towards $v_{P}$. Yet this point is unstable and, unless we have started directly from the separatrix S, the magnitude of the envelope ultimately diverges. In the case $|\eta|/\Gamma>\sqrt{2}$, the fixed point $P$ has disappeared and the dynamics again either blows up if the initial condition sits above the separatrix or dies down if the initial condition sits below (Fig.~\ref{Fig4}c).

There is hence a saddle-node bifurcation at $|\eta|/\Gamma=\sqrt{2}$ at which the fixed point $P$ disappears (Fig.~\ref{Fig4}d). The soliton never persists because its dynamics is controlled by an unstable fixed point except precisely at this bifurcation, where the timescale---hence the lifetime of the soliton---diverges.

To confirm these findings, we solve Eqs.~(\ref{ffu}-\ref{ffv}) numerically and find the two regimes reported above depending on the ratio between non-reciprocity and dissipation $\eta/\Gamma$ (Fig.~\ref{Fig4}ef) and on the magnitude of the initial condition (Fig.~\ref{Fig4}gh):  (i) above the separatrix  $\eta/\Gamma =1.50$, the soliton accelerates, decays in a short transient before blowing up (red curve);  (ii) below the separatrix $\eta/\Gamma =1.00$, the soliton accelerates until a terminal velocity while its magnitude decreases (blue curve); (iii) finally, close to the separatrix  $\eta/\Gamma =1.373$, the soliton accelerates, its magnitude decreases, but after a transient blows up (green curve). The closer to the separatrix and the closer to the bifurcation, the longer the transient. We further confirm these results with numerical solutions to Eq.~\eqref{nls0}. The results are consistent even though additional radiations in Eq.~\eqref{nls0} induce a quantitative discrepancy with the solutions of Eqs.~(\ref{ffu}-\ref{ffv}). Such discrepancy is natural given that these equations describe a single breathing soliton, in the absence of additional waves such as radiations (see Appendix ~\eqref{nseP} for a discussion on the role of such radiations).

How do these predictions compare to the numerical solutions of Eq.~\ref{sineG}?  We readily see that the ratio $\eta/\Gamma$ for which the longest living breather is $\eta/\Gamma\simeq 1.4$ is consistent with the prediction $\eta/\Gamma=\sqrt{2}$ above. The theory also predicts a breather velocity of $\sqrt{2}$, which is fully consistent with the simulation shown in Fig.~\ref{Fig3}a~\footnote{The breather velocity in the numerical solution to Eq.~\eqref{sineG} was $\sqrt{2}/2$, which is consistent with the $\sqrt{2}$ prediction given the change of variable $t \to t/2$ done to derive Eq.~\eqref{nls0}}. What about the initial condition? Given that the initial conditions used to numerically solve Eq.~\eqref{sineG} are that of a static breather (Eq.~\eqref{approx22}), the correspondence between the parameters $\Pi$, $d$, $u(0)$ and $v(0)$ is the following: $u(0)=(\Pi d -2)/(2d)$ and $v(0)=0$. Therefore the optimal initial condition to sit on the separatrix is $u(0)=u_S=\sqrt{2}\sqrt{1-(\eta/\Gamma)^2/5}$, which corresponds to $\Pi_S=(2+2 d \sqrt{2}\sqrt{1-(\eta/\Gamma)^2/5})/d=2 \sqrt{6/5}+2\simeq4.19$ in the limit case $\eta/\Gamma=\sqrt{2}$---we chose $d=1$ for our simulations. This prediction is in beautiful agreement with our simulations, where we found that the optimal initial conditions to observe the longest breathers were $\Pi=4.2$.

In conclusion, we now understand why the region of existence of the breather is so narrow: its dynamics is underpinned by an unstable fixed point, which leads to a long-lived soliton if two conditions are met: (i) while this fixed point exists for a wide range of parameters, only when the ratio between non-reciprocity and dissipation is $\eta/\Gamma=\pm\sqrt{2}$ will this pair of fixed points undergo a bifurcation and the timescale of the transient diverge; (ii) only when the initial conditions are precisely tuned close to the separatrix will the dynamics approach this fixed point.

\indent These results are seemingly at odds with the experimental and numerical observations reported in Fig.~\ref{Fig2}, where long lasting breathing solitons were also observed for a wide range of initial conditions. To resolve this issue, we run simulations of Eq.~\eqref{eq:NRFK} with varying degrees of discreteness $D$ (See Appendix ~\ref{Experimental Methods} and Fig.~\ref{transition}). We observe that as the discreteness parameter $D$ decreases, viz. as we tend towards the continuum limit, the region of stability progressively shrinks and converges towards a point. Hence, in the discrete case, the line of stability appears to be a vestige of the island of stability we predicted analytically above in the section, but discreteness enhances stability.

\section{Effect of Discreteness}

There is a powerful way to address the role of discreteness in the NLS model. Indeed, contrary to the FK model, the NLS equation \eqref{nls0} admits an integrable discretization known as the Ablowitz-Ladik (AL) model \cite{ablowitz1974inverse,ablowitz1975nonlinear,ablowitz1976nonlinear}. Following the ideas presented in \cite{Kivshar_Campbell}, we therefore focus on the discrete form of \eqref{nls0}, given by
\begin{multline}\label{AL}
    i\partial_t \psi_n + \frac{\psi_{n+1}+\psi_{n-1}-2\psi_n}{a^2}   + |\psi_n|^2 (\psi_{n+1}+\psi_{n-1})\\ = R_n[\psi] + R^{(d)}_n[\psi].
\end{multline}
Here, the left-hand side represents the aforementioned AL discretization, with $a$ denoting the lattice spacing. This system possesses an exact one-soliton solution, parametrized by constants $u$ and $v$, of the form
\begin{equation}\label{1solD}
\psi^{(s)}_n(t) = \frac{\sinh (  a u ) e^{i v n a +i \Omega(t)}}{a \cosh(u(n a -x_c(t)))},
\end{equation}
where the soliton center and phase evolve as
\begin{equation}
\frac{dx_c}{dt} = \frac{2\sin(a v)\sinh(a u)}{a^2 u},
\end{equation}
\begin{equation}
\frac{d\Omega}{dt} = \frac{2(\cos(av )\cosh(a u)-1)}{a^2}.
\end{equation}
In the continuum limit $a \to 0$, this discrete solution converges to the continuous profile \eqref{profS}.
The perturbation terms $R_n(\psi)$ are defined by the physical damping and non-reciprocity 
\begin{equation}
    R_n[\psi] = \eta \frac{\psi_{n+1}-\psi_n}{a} - i \Gamma \psi_n+ \frac{\Gamma}{2}\partial_t \psi_n
\end{equation}
while the terms $R^{(d)}[\psi]$ measure the difference between the naive discretization of the NLS model and the AL discretization \cite{Kivshar_Campbell}
\begin{equation}
    R^{(d)}_n[\psi] = |\psi_n|^2 (\psi_{n+1}+\psi_{n-1}-2\psi_n).
\end{equation}
The evolution of the soliton parameters $u$ and $v$ can be derived using inverse scattering perturbation theory, as in the continuous case (see also \cite{Vakhnenko1986}), or alternatively through the analysis of conserved quantities. This yields
\begin{equation}\label{uD}
   \frac{du}{dt} =X^{(u)}_a(u,v), 
\end{equation}
\begin{equation}\label{vD}
    \frac{dv}{dt} =X^{(v)}_a(u,v)  - \frac{(2\pi)^3 \sinh(a u)^2}{a^6u^3} \sum\limits_{n=1}^\infty \frac{n^2\sin(2\pi n x_c/a)}{\sinh \frac{\pi^2 n }{ u a}},
\end{equation}
with the explicit expressions for the vector fields $X^{(u,v)}_a$ provided in the Appendix (Eqs. \eqref{uD1}, \eqref{vD1}).

\begin{figure}[t!]
\centering
\hspace{0in}
\includegraphics[width=1.0\columnwidth,trim=0cm 0cm 0cm 0cm]{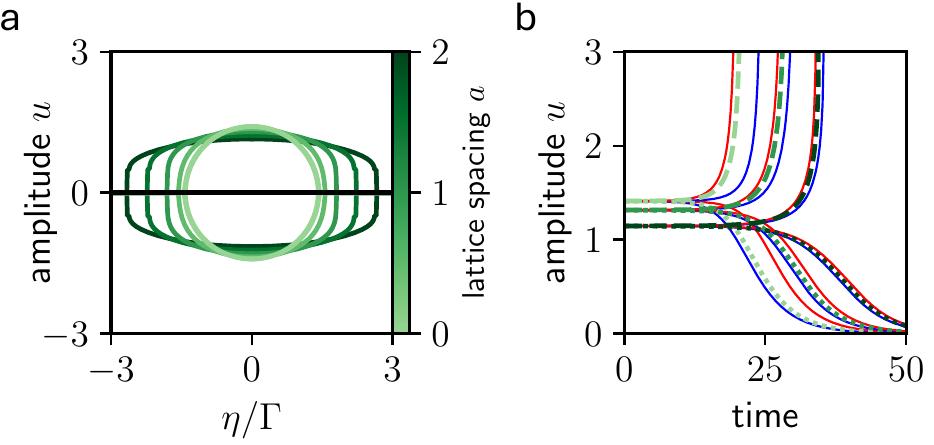}
\caption{\textbf{Discrete effects in the nonlinear Schr\"odinger equation.}
{\bf (a)} Bifurcation diagram as a function of the lattice spacing $a$. The lightgreen line corresponds to the fixed points depicted in Fig.~\ref{Fig4}d.
% {\bf (b)} Bifurcation point $|\eta_c|/\Gamma$ as a function of the lattice spacing $a$.
{\bf (b)} Time evolution of the amplitude according to Eqs.~\eqref{uD}-~\eqref{vD}, with initial conditions ($u$,$v$) corresponding to the system's fixed point but with the $u$-coordinate perturbed in the positive (negative) direction by $0.3\%$, shown in the dashed (dotted) lines, for three lattice spacings $a=[0.1,1,2]$ and with $\eta=0.01$ and $\Gamma=0.1$. Red and blue lines surrounding the green lines indicate the effect of an additional perturbation in $\eta$ by $\pm0.015\%$.
% {\bf (d)} 
}
\label{Fig5}
\end{figure}

In the limit $a \to 0$, these expressions reduce to the continuum dynamics \eqref{ffu}, \eqref{ffv}.
The final term in \eqref{vD} represents the Peierls-Nabarro (PN) contribution, arising from the lattice structure. In our analysis, we retain only the $n=1$ term in this sum, an approximation also performed in~\cite{Kivshar_Campbell}, and neglect the corresponding PN contributions to $X_a^{(u,v)}$, as they are small not only due to discreteness, but also due to the perturbations. 
This way, $X_a^{(u,v)}$ define again a two-dimensional flow, qualitatively similar to that shown in Fig.~\ref{Fig4}abc. However, the bifurcation point now depends on the lattice spacing through the modified condition $|\eta|/\Gamma \le \sqrt{2 + a^2}$, which flattens the bifurcation line of Fig.~\ref{Fig4}d (see Fig.~\ref{Fig5}a). Hence, the sensitivity to the magnitude of the initial condition is weaker, in accordance with our findings of Fig.~\ref{transition}. 
To better isolate the role of lattice spacing, we focus on the regime of small $\eta$, where the dependence of the bifurcation point on $a$ is minimal. It is important to note that the inclusion of the PN term distorts the simple 2D flow structure. We therefore compute the critical points numerically and analyze the perturbations around them. The corresponding results shown in Fig.~\ref{Fig5}b demonstrate that discreteness enhances soliton lifetime but also renders the dynamics less sensitive to changes in the non-reciprocal parameter $\eta$.

\section{Discussion}

The existence of these genuine breathing solitons in non-reciprocal nonlinear media is in stark contrast with that of solitons in constantly driven media, which admit static short-lived breathers~\cite{Lucas_NatComm2017,Yu_NatComm2017,Pernet_NatPhys2022}, yet whose behavior does not seem to correspond to any fixed point for reasonable values of the driving (see Appendix \eqref{NLSconstant}).
We emphasize here that our considerations for the soliton solution in the non-reciprocal nonlinear Schr\"odinger Equation~\eqref{nls0} serve as an enveloping profile for the  breather of the non-reciprocal sine-Gordon Equation~\eqref{sineG}. In contrast, the breathers in~\cite{Lucas_NatComm2017,Yu_NatComm2017,Pernet_NatPhys2022} from the point of view of the driven nonlinear Schr\"odinger Equation correspond to the special solutions of Kuznetsov-Ma type~\cite{Kuznetsov1977,Ma1979} that could be obtained by a dressing or degeneration of the periodic solutions~\cite{Ma1979}. In addition, other types of static breathing solitons, sometimes called oscillons~\cite{Oscillons_JHEP2024}, have a more complicated solution to the nonlinear sine-Gordon equation called Dashen, Hasslacher and Neveu (DHN) solution for which an inverse scattering analysis is impossible. Crucially, both these types of standing breathers are static, in contrast to our non-reciprocal breathers that travel.

We interpret the difference between these types of breathers as follows. A breathing soliton can be seen as a bound pair of kink and antikink that oscillates in time. In ordinary driven media, kinks and antikinks are driven towards opposite directions~\cite{dauxois2006physics}. This means that the driving tends to keep the breather from traveling and to pull the kink and antikink apart, thereby to disintegrate the breather. In contrast, in non-reciprocal media, kinks and antikinks are driven towards the same direction~\cite{Veenstra_Nature2024}. This means that the non-reciprocal driving has the tendency to drive the breather unidirectionally and to maintain its shape. Our results here have demonstrated that such kink-antikink pair making up the breather is however a delicate balance and only persists if the energy injection precisely compensates the dissipation and if the initial conditions have triggered a breather with the suitable magnitude.

In conclusion, we have discovered how non-reciprocity affects the emergence and long time dynamics of breathing solitons in a non-reciprocal active material. Our exploration has been performed experimentally, numerically, and theoretically for non-reciprocal generalizations of the sine-Gordon and nonlinear Schr\"odinger equations.
In the generic case, non-reciprocity amplifies localized initial conditions towards unidirectional nonlinear structures such as radiations, single or multiple breathers, or kink-antikink pairs. Such nonlinear excitations generically either are damped or diverge over short timescales. Only for a specific region of parameter space, when non-reciprocal gain, dissipation and initial magnitude are carefully balanced, will a pure breathing soliton emerge, propagate unidirectionally and remain stable over a long timescale. This careful balance correspond to the bifurcation of a pair of unstable fixed points, which can be reached by the dynamics provided the initial conditions are precisely tuned. 
In addition to these continuum theoretical results
we have extended our analysis to a discrete setting, where we found that discreteness has a stabilizing effect and tends to expand the regime of existence and lifetime of non-reciprocal breathers, which explains our experimental and numerical observations.

We have identified a minimal set of ingredients necessary for non-reciprocal breathers to emerge: weakly nonlinear effects must stabilize otherwise dispersive waves while non-reciprocal forces counteract dissipation to sustain breather solitons. 
Non-reciprocity itself may arise in a variety of ways. 
In the current work, these interactions are explicitly programmed and applied using electrical power source,
but they arise generically in systems with asymmetric phase delays~\cite{DelPino_Nature2022,Wanjura_NatPhys2023}, asymmetric exchange interactions~\cite{Hanai2024-jw}, mediated by hydrodynamic fields~\cite{Beatus2006-xn,Bililign2021Motile, Poncet_PRL2022,TanNature,Guillet2025-ap}, acoustic streaming ~\cite{Lim2019NatPhys,jaegernonreciprocity2025} or light fields~\cite{ZheludevNatPhys,delic2024}.
In the majority of these systems, mechanical oscillations and waves can exist and hence these platforms are candidates to potentially host non-reciprocal breathers.

Our work establishes non-reciprocity as a mechanism of choice to obtain traveling breathing solitons in active/dissipative materials and hence to transport information and energy in dissipative media.
Open questions ahead are how to increase further the range of stability of such breathers and to explore their stability to disorder and noise.

\textit{Acknowledgments.} --- We thank Daan Giesen and Tjeerd Weijers for technical support.
We thank Filip Novkoski and
Stephane Randoux for discussion on the numerical implementation of the inverse scattering method.
We acknowledge funding from the European Research Council under grant agreement 852587 and from the Netherlands Organisation for Scientific Research under grant agreement VI.Vidi.213.131. M.B. acknowledges funding from the European Research Council under Grant Agreement No. 101117080. The data and codes supporting this study are freely available~\cite{gitlabrepo}.
% \url{https://uva-hva.gitlab.host/published-projects/non-reciprocal-breathing-solitons.git}.

\bibliography{refs.bib}

\clearpage

\section{Experimental Methods}\label{Experimental Methods}

\begin{figure*}[t!]
\centering
\hspace{0in}
\includegraphics[width=2\columnwidth]{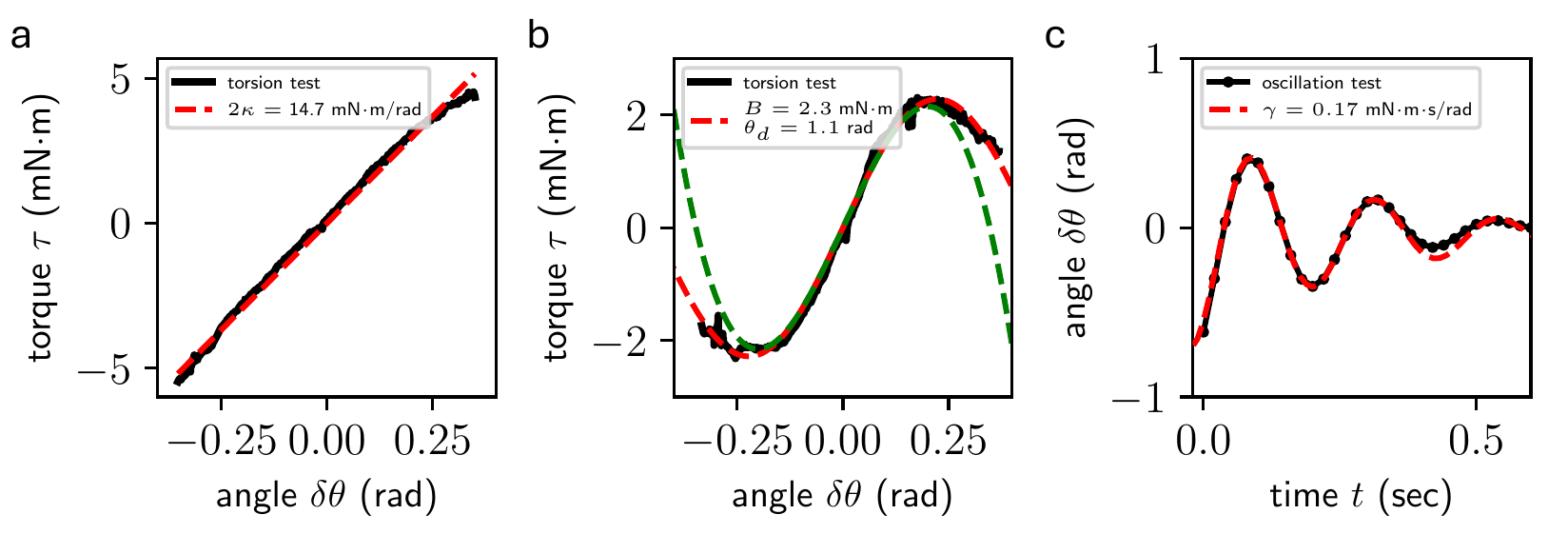}
\caption{\textbf{Calibration of experimental model parameters.}
{\bf (a)} Torque-angle curve measuring the elastic response of a single unit without the magnetic and non-reciprocal contributions. While keeping the angles of the two neighboring oscillators fixed, the elastic torque generated is measured with an Instron E3000 linear-torsional testing machine. A linear fit yields $\kappa \approx 7.5 \ \mathrm{mN}\cdot\mathrm{m}$/rad.
{\bf (b)} Torque-angle curve measuring the magnetic response of a single unit without the elastic and non-reciprocal contributions. A sinusoidal fit yields amplitude $B \approx 2.3 \ \mathrm{mN}$ and period $\theta_d \approx 1$rad. The dashed green line indicates the third order Taylor expansion of the fitted sine function, which validates the weakly nonlinear approximation done in the Main Text. 
{\bf (c)} Decaying oscillation after perturbation of a single oscillator. We fit the data to a decaying sine function $\propto \exp( -\gamma t) \cos(\omega t)$ to recover the damping coefficient $\gamma = 0.17$ mN$\cdot$m$\cdot$s/rad.}
\label{Fig:calibration}
\end{figure*}

Fig. \ref{Fig1}a shows four coupled oscillators that are part of an active metamaterial consisting of 50 identical unit cells, separated by a lattice spacing of $a=6$cm. 
The unit cells are directly inspired by \cite{Veenstra_Nature2024}.
Each oscillator arm has a rotational degree of freedom $\theta_i$ with moment of inertia $I = 3 \cdot 10^{-5}$ kg$\cdot$m$^2$ and is coupled elastically to its neighbors by 2 rubber bands of length 4.6 cm and thickness 2 mm. 
As the difference in angle between two neighbors $\theta_i - \theta_{i+1}$ changes, the rubber bands induce a torque $\tau_{i} = \kappa (\theta_i - \theta_{i+1})$ where $\kappa = 7.5 \ \mathrm{mN}\cdot\mathrm{m}$/rad is the torsional stiffness that we calibrate using a torsion testing machine (Instron E3000 linear-torsional, see Fig.~\ref{Fig:calibration}a). Besides these passive elastic interactions that are symmetric with respect to left and right neighboring sites $i-1$ and $i+1$, each unit incorporates an electromechanical motor implementing an active feedback that breaks this symmetry. Each unit applies a torque
\begin{equation}
\tau_i^a= \kappa^a (\theta_{i-1}- \theta_{i+1}),
\label{eq:feedback}
\end{equation}
where $\kappa^a$ represents the non-reciprocal interaction strength. Angular positions are measured and recorded by an angular encoder (Broad-com HEDR-55L2-BY09) present in each unit cell. This information is transmitted between nearest neighbors at a rate of $100$ Hz and used to compute the torque $\tau_i^a$ to be exerted by a direct-current coreless motor (Motraxx CL1628) via a microcontroller (Espressif ESP 32), both of which are embedded in the unit cell.
Each oscillator arm holds a neodymium magnet which interacts with an identical magnet placed regularly on a substrate. Contrary to the setup of \cite{Veenstra_Nature2024}, where substrate magnets are spaced alternatingly in between oscillators, here we align magnets directly opposite each oscillator arm. 
This generates a weakly nonlinear monostable magnetic potential that induces a torque that we measure to be approximately sinusoidal with the oscillator angle $\theta_i$ and has an amplitude of $B=2.3$ $\mathrm{mN}\cdot\mathrm{m}$ (see Fig.~\ref{Fig:calibration}b).
The viscous dissipation associated to oscillator rotation was found to be $\gamma = 0.17 $ m$\cdot$Nm$\cdot$s \ rad by fitting the oscillation amplitude decay after an initial perturbation (see Fig. \ref{Fig:calibration}c).

For an appropriate non-reciprocal interaction strength $\kappa^a$, the perturbation sets in motion the breathing pulse shown in Fig.\ref{Fig1}c. 
Using the total squared amplitude as a measure for the system's energy $E=\sum_i \theta_i^2$, we quantify the breather's lifetime by measuring the ratio between minimum and maximum energy within a time window of one starting shortly after the initial kick has ringed down. This ratio will be close to $1$ for waves whose energy does not change significantly over time, while it vanishes for decaying and amplifying waves.
In Fig.\ref{Fig2}b, we plot the logarithm of this quantity for a sweep over initial velocities and non-reciprocal coupling strengths. Note that alternative choices for the time window only change the the scale of the colormap but preserve the observed bifurcation between decaying and amplifying waves.

\section{Numerical Methods}\label{Numerical Methods}
\subsection{Non-reciprocal Frenkel-Kontorova model}

We model the dominant forces -- inertial, elastic, viscous, non-reciprocal and magnetic -- present in the active metamaterial with a non-reciprocal version of the Frenkel-Kontorova model:
\begin{equation}
\begin{split}
I \pdv[2]{\theta_i}{\tau} \!= & \kappa (\theta_{i-1}\!+\!\theta_{i+1}\!-\!2\theta_{i})\!-\!\kappa^a (\theta_{i+1}\! -\! \theta_{i-1}) \!\\& -\! \gamma \pdv{\theta_i}{\tau} \!+\! B \sin(2\pi\frac{\theta_i}{\theta_d} )
\end{split}
\label{eq:NRFKdim}
\end{equation}
where $\tau$ denotes dimensionful time and $\theta_d=1$ rad is the width of the magnetic potential, found by fitting the measured magnetic field to a sine function (see Fig.~\ref{Fig:calibration}b).
By employing the following substitutions, we find the non-dimensional form of Eq.~\eqref{eq:NRFK}:
\begin{equation}
\begin{aligned}
    \phi_i &= 2\pi\frac{\theta_i }{\theta_d}\\
    t &= \sqrt{\frac{\kappa}{I}} \tau \\  
    \eta &= \frac{2\kappa^a}{ \kappa \sqrt{D}}\\
    \Gamma &= \frac{\gamma}{\sqrt{I \kappa D}} \\
    D &= \frac{2\pi B}{\kappa\theta_d} \\
    \label{parameters}
\end{aligned}
\end{equation}
We note that this model is identical to the one discussed in ~\cite{Veenstra_Nature2024} in the context of topological solitons but we focus here instead on the weakly nonlinear regime where oscillators do not hop between different minima. 
To compare model to experiment, we integrate Eq.~\eqref{eq:NRFK} with the velocity Verlet algorithm for a system of $N=50$ oscillators using the experimentally calibrated parameters and timestep $dt=0.1$. Mirroring the experiment, in Fig.~\ref{Fig2}cd, a single site is initialized with a nonzero angular velocity $\dot{\theta}_{t=0}$ and stress free boundary conditions are imposed.

\begin{figure}[t!]
%\hspace{-2.2cm}
\centering
\includegraphics[width=1.0\columnwidth,trim=0cm 0cm 0cm 0cm]{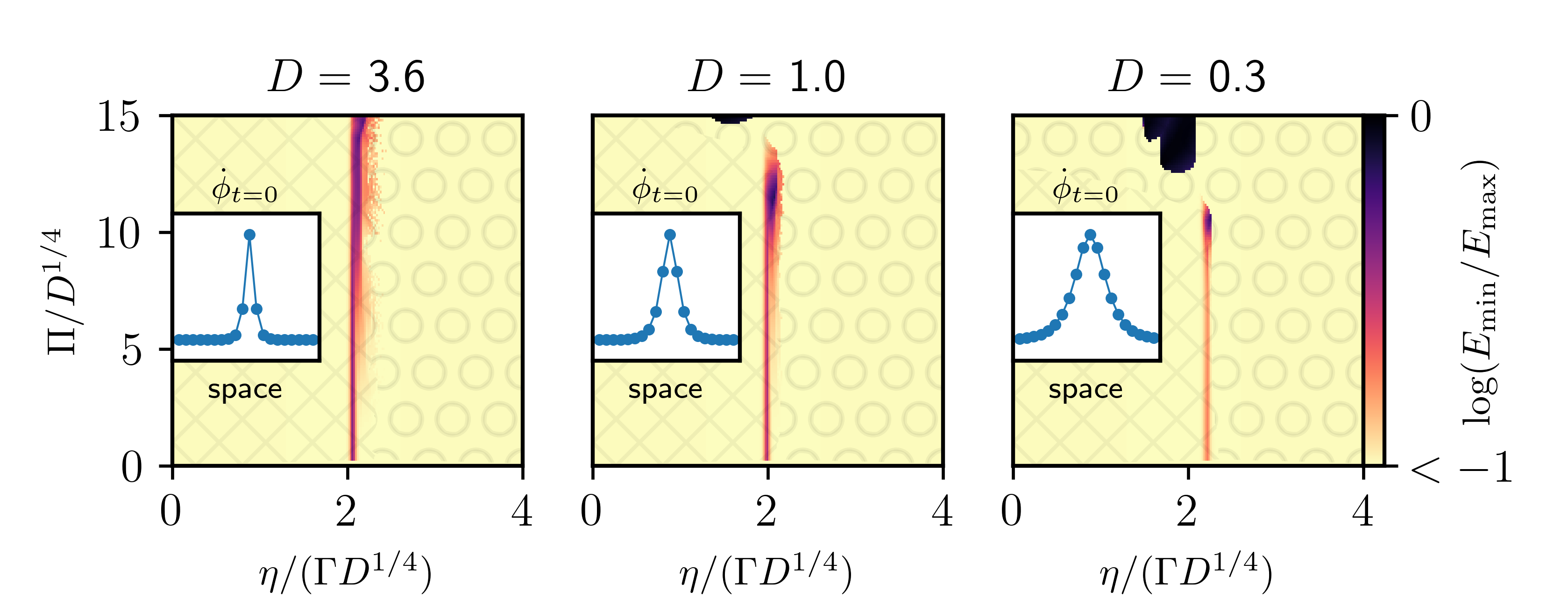}
\caption{\textbf{Stability of breathers with decreasing discreteness.}
Phase diagram  of the logarithm of the ratio between minimum and maximum energy $\sum \theta^2_i$ visited by the system up until non-dimensional time $t=1000$. As $D$ decreases, the vertical stable region also seen in Fig.~\ref{Fig2}bd morphs into a pointlike region as seen and predicted in Fig.~\ref{Fig3}c. The black region in the $D=0.3$ case corresponds to topological solitons, also observed in Fig.~\ref{Fig3}c.
Regions hatched with squares (circles) demarcate the weakly (strongly) nonlinear regime of Eq. (2).  Insets show the shape of the initial condition $\dot{\phi}_{t=0}$.}
\label{transition}
\end{figure}

In Fig~\ref{transition}, we plot the breather stability for $D=\{3.6,1.0,0.3\}$ and observe that as $D$ decreases and the continuum limit is approached, the stable region depends not only on the activity $\eta$, but also on the amplitude of the initial condition $\Pi$. This confirms our finding that breathers in the continuum descriptions discussed in Figs.~\ref{Fig3} and ~\ref{Fig4} of the Main Text, are stable only when both activity and initial condition are tuned. These findings also suggest that discreteness enhances breather stability in non-reciprocal systems.

\subsection{Simulation of continuum equations}

Simulations of both the non-reciprocal sine-Gordon and nonlinear Schr\"odinger equations, shown in Fig \ref{Fig3} and Fig \ref{Fig4} respectively, were performed using the open source package \textit{py-pde} \cite{py-pde} with adaptive timestep and 601 spatial discretization steps.
For Fig \ref{Fig3}bc, the initial condition corresponds to the breather profile described by eq. \eqref{sGprofile}, while Fig \ref{Fig4} uses eq. \eqref{profS} as an initial condition with $\Phi(0)=0$.

\section{Theory}

\subsection{Continuum limit of the discrete model}
\label{contlim}
The continuum limit of equation~\eqref{eq:NRFK} is found by letting $\phi_i$ become a continuous function $\phi(x)$ of space $x \in [0, Na]$, with N the number of units. 
We approximate finite differences by a Taylor expansion according to
%\begin{equation}
${    \phi_{i+1}-\phi_i \approx   a \phi_x + a^2\phi_{xx}/2}$
%\end{equation}
and substituting terms in the discrete model of Eq.~\eqref{eq:NRFK} then leads to Eq.~\eqref{sineG} under rescaling of the spatial variable $x \rightarrow \frac{a}{\sqrt{D}}x$ and time $t\rightarrow \frac{t}{\sqrt{D}}$.

\subsection{Inverse scattering method for the sine-Gordon equation}
\label{theor1}

In this chapter we briefly review the main points of the inverse scattering method for solving
classical integrable $(1+1)$-dimensional field theories. We follow mainly \cite{Faddeev_1987}.
We start with the sine-Gordon equation
	\begin{equation}\label{sg1}
		\partial_t^2 \varphi - \partial_x^2 \varphi + \sin \varphi = 0.
	\end{equation}
The key observation is that this equation is equivalent to the following \textit{auxiliary} linear system
\begin{equation}\label{sys1}
	\frac{\partial F}{\partial x} = U F ,\qquad \frac{\partial F}{\partial t} = V F
\end{equation}
where $U$ and $V$ are $2\times 2$ matrices that depend on the field $\varphi$ and the auxiliary spectral parameter $\lambda$
\begin{equation}
	U =\frac{\partial_t \varphi \sigma_3 }{4i} + \frac{\lambda+\lambda^{-1}}{4i} \sigma_1\sin \frac{\varphi }{2}
	+ \frac{\lambda-\lambda^{-1}}{4i} \sigma_2\cos \frac{\varphi }{2}\end{equation}
	\begin{equation}
	V=
	\frac{\partial_x \varphi \sigma_3}{4i}  + \frac{\lambda-\lambda^{-1}}{4i} \sigma_1\sin \frac{ \varphi}{2}
	+ \frac{\lambda+\lambda^{-1}}{4i} \sigma_2\cos \frac{\varphi }{2}
\end{equation}
The equivalence to Eq. \eqref{sg1} comes from the compatibility condition of these linear systems (also called zero-curvature condition)
\begin{equation}
	\frac{\partial}{\partial x} \frac{\partial}{\partial t} F = 	\frac{\partial}{\partial t} \frac{\partial}{\partial x} F
\end{equation}
or equivalently
\begin{equation}\label{zeroC}
	\partial_t U - \partial_x V + [U,V] =0.
\end{equation}
Demanding that this relation is satisfied for every value of $\lambda$ we obtain our original equation \eqref{sg1}.

The important part of the scattering formalism are Jost solutions $T_{\pm}(x,\lambda)$. They are defined as solutions of the linear system \eqref{sys1} at the fixed moment of time
\begin{equation}\label{jost2}
	\frac{d T_{\pm}(x,\lambda)}{dx} = U T_{\pm}(x,\lambda)
\end{equation}
and are specified by the following asymptotic behaviour

\begin{equation}\label{aa1}
	T_{-}(x\to -\infty) = \mathcal{E} e^{-ix \sigma_3 (\lambda^2-1)/(4\lambda)}\equiv E(x,\lambda)
\end{equation}
\begin{equation}\label{aa2}
	T_+(x\to +\infty) = \begin{cases}
		(-1)^{Q/2} E(x,\lambda), &  Q  \quad {\rm even}  \\
		(-1)^{(Q-1)/2} i\sigma_3 E(x,\lambda) ,& Q  \quad {\rm odd}
	\end{cases}
\end{equation}
where matrix $\mathcal{E}$ and topological charge $Q$ are given by
\begin{equation}
	\mathcal{E} = \frac{1}{\sqrt{2}} \left(
	\begin{array}{cc}
		1 & i \\
		i & 1
	\end{array}
	\right),\quad Q =   \frac{1}{2\pi} \int\limits_{-\infty}^\infty dx \partial_x \varphi(x)
\end{equation}
The reason behind these definitions is twofold: first, we consider very specific solutions of the sine-Gordon equation, that at $x\to \pm \infty$ approach constant value
$\partial_t\varphi(x\to \pm \infty) = 0$, moreover these values can differ by the topological charge
\begin{equation}\label{q}
	\varphi(+\infty) - \varphi(-\infty)  = \frac{1}{2\pi} \int dx \partial_x \varphi = Q .
\end{equation}
In particular, we can put $\varphi(-\infty)=0$ so the function $E(x,\lambda)$ is a solution of the linear problem with the asymptotic matrix $U(x\to-\infty)$.
The second reason is more subtle since obviously the solutions of the linear problem \eqref{jost2}  are defined up to multiplication on the constant matrix from the right.
The choice of this matrix that leads to the definition \eqref{aa1} is dictated by the analytic properties of the Jost solutions as functions of the spectral parameter $\lambda$, namely, one can show that similar to the columns of $E(x,\lambda)$ the columns $T^{(l)}_\pm$, $l=1,2$
of the Jost matrices $T_\pm(x,\lambda)$ have the following analytical properties: $T_-^{(1)}$ and $T_+^{(2)}$ can be analytically continued to the upper half plane, and
$T_-^{(2)}$ and $T_+^{(1)}$ to the lower half plane of the complex variable $\lambda$. Moreover, the specific form of the differential equation allows for the following conjugational properties (involutions)
\begin{equation}\label{invo}
	\bar{T}_\pm (x,\lambda) = \sigma_2 T_{\pm}(x,\lambda) \sigma_2 ,\qquad
	\bar{T}_\pm (x,-\lambda) = -i\sigma_1 T_{\pm}(x,\lambda).
\end{equation}

The Jost solutions should be related by the constant matrix multiplied from the right. Traditionally, this matrix is called the transfer matrix
\begin{equation}\label{ttSG}
	T_- (x,\lambda) = T_+(x,\lambda) T(\lambda),\quad T(\lambda) = \left(
	\begin{array}{cc}
		a(\lambda) & - \bar{b}(\lambda) \\
		b(\lambda) & \bar{a}(\lambda)
	\end{array}
	\right).
\end{equation}
Since the matrix $U$ is traceless the determinant of Jost solutions does not change with time $\partial_t \det T(x) = 0$, which means the unimodularity of the transfer matrix $T(\lambda)$
\begin{equation}\label{detTT}
	\det T(\lambda) = |a(\lambda)|^2 + |b(\lambda)|^2 =1.
\end{equation}
The form of the transfer matrix $T(\lambda)$ follows from the first of the involutions \eqref{invo}, which implies
\begin{equation}
	\bar{T}(\lambda) = \sigma_2 T(\lambda) \sigma_2.
\end{equation}
The second involution implies
\begin{equation}
	\bar{T}(\lambda) = T(-\lambda) ,
\end{equation}
which is equivalent
\begin{equation}\label{aa11}
	a(-\lambda) = \bar{a}(\lambda),\qquad
	b(-\lambda) = \bar{b}(\lambda).
\end{equation}
Moreover, we can express
\begin{equation}\label{deta!!}
	a(\lambda) = \det(T_-^{(1)},T_+^{(2)} ),
\end{equation}
therefore $a(\lambda)$ can be analytically continued to the upper half plane.
The bound states in the scattering are described by zeroes in the upper half plane, which due to Eq. \eqref{aa11} can be either purely imaginary $\lambda = i\kappa_n$, $\kappa_n>0$, or come in pairs $\lambda = \lambda_n$, $\lambda = -\bar\lambda_n$, ${\rm Im} \lambda_n >0$ (see \eqref{invo}).
Moreover, using Kramers–Kronig relations along with Eq. \eqref{detTT} one can present
\begin{equation}\label{apres}
	a(\lambda) = R(\lambda)\prod\limits_{n=1}^{N_1} \frac{\lambda - i \kappa_n}{\lambda + i \kappa_n } \prod\limits_{k=1}^{N_2} \frac{\lambda - \lambda_k }{\lambda - \bar\lambda_k}
	\frac{\lambda +\bar\lambda_k }{\lambda + \lambda_k}
\end{equation}
with $R(\lambda)$ is referred in the main text as the radiative part and is given explicitly as
\begin{equation}
    R(\lambda) =\exp \left(\int \frac{\ln (1-|b(\mu)|^2)}{\mu - \lambda-i0}\frac{d\mu}{2\pi i}\right).
\end{equation}
From the presentation \eqref{deta!!} we see that whenever the spectral parameter coincides with  one the zeroes of  $a(\lambda)$, $\lambda=\lambda_j$, - there is proportionality between
between the columns
\begin{equation}\label{gammaSG}
	T_-^{(1)}(x,\lambda_j) =\mathfrak{g}_j T_+^{(2)}(x,\lambda_j) ,\qquad a(\lambda_j) =0.
\end{equation}
The proportionality coefficient $\mathfrak{g}_j$ is another scattering data along with the coefficient $b(\lambda)$ and the position of zeroes $\lambda_j$.
The time evolution of the scattering data can be deduced from the second equation in \eqref{sys1} or the zero curvature condition \eqref{zeroC}

\begin{align}\label{dyn1SG}
	a(\lambda,t) &= a(\lambda,0),\qquad b(\lambda,t)= b(\lambda,0) e^{it(\lambda^2+1)/(2\lambda) }, \\ \mathfrak{g}_j(t) & = e^{it(\lambda_j^2+1)/(2\lambda_j) } \mathfrak{g}_j(0).
\end{align}

Notice that the nonlinear time dynamics of the original equation reduces to the extremely simple evolution on the scattering data. Therefore, the typical way to solve the Cauchy problem for the sine-Gordon equation is to first find the scattering data i.e. to solve \eqref{jost2} (so-called direct scattering problem).
Then evolve it according to \eqref{dyn1SG}, afterwards restore the potential by means of the inverse scattering problem i.e. from the given scattering data.
The last step involves in general solving linear integral equations to recover the profile, but for the so-called reflectionless potentials, for which $b(\lambda)\equiv 0$ these integral equations transform into the algebraic ones, and the corresponding profiles have
profound physical meaning.
In particular, for
\begin{equation}
    a(\lambda) = \frac{\lambda - \lambda_1}{\lambda-\bar\lambda_1}\frac{\lambda + \bar\lambda_1}{\lambda+\lambda_1},\quad \lambda_1 = \xi + i \nu
\end{equation}
The corresponding profile is given by a breather (see Eq.~\eqref{phi0}).
A solution that corresponds to a single imaginary zero
\begin{equation}
    a(\lambda) = \frac{\lambda- i \kappa }{\lambda+ i \kappa } ,\qquad \kappa>0
\end{equation}
is called a kink (soliton), and has a profile
\begin{equation}\label{tempF}
	\varphi(x,t) = 4 \arctan \exp\left(\frac{x- v t - x_0}{\sqrt{1-v^2}}\right).
\end{equation}
Contrary to a breather this solution has a non-zero topological charge $Q = +1$ (see \eqref{q}).
The antikink is obtained by changing the overall sign in \eqref{tempF} and corresponds to $Q=-1$.

As an example of the direct scattering problem, let us discuss a spectral content of given by the following profile
\begin{equation}
	\varphi(x,0) = 0 , \qquad \partial_t \varphi(x,0) = \frac{\Pi}{\cosh(x/d)}.
\end{equation}
This profile is used to imitate the initial kick in the experiment.
To compute the scattering data we make a substitution $F = \mathcal{E} f$ in the linear system \eqref{sys1},
rescale $x = 4y/\Pi$ and denote $p=2(\lambda-1/\lambda)/\Pi$,
which leads to
\begin{equation}
	\frac{d f}{dy} = \left(\frac{\sigma_3p}{2i} -\frac{i \sigma_2}{\cosh \frac{4y}{\Pi d}}\right)f
\end{equation}
Taking into account results from \cite{Gamayun2016} we obtain
\begin{equation}\label{scat1}
	a(\lambda) = \frac{\Gamma^2\left(\frac{1}{2} - i \frac{(\lambda-1/\lambda)d}{4}\right)}{\Gamma\left(\frac{1}{2} - i \frac{(\lambda-1/\lambda)d}{4} - \frac{\Pi d}{4}\right)\Gamma\left(\frac{1}{2} - i \frac{(\lambda-1/\lambda)d}{4} + \frac{\Pi d}{4}\right)}
\end{equation}
\begin{equation}\label{scat2}
	b(\lambda) = \frac{\sin\frac{\pi\Pi d}{4}  }{\cosh \frac{\pi(\lambda-1/\lambda) \Pi d}{8}}
\end{equation}
So $b(\lambda)=0$ for $\Pi d = 4 n$, $n \in \mathds{Z}_+$. For instance, $n=1$ correspond to
\begin{equation}
	a(\lambda) = \frac{\lambda^2 - 2i \lambda /d -1}{\lambda^2 + 2i \lambda /d -1}
\end{equation}
So for $d<1$ we have kink-antikink pair and for $d>1$ we have breather.
The special case $\Pi  =4$, $d=1$ implies the ``degenerate'' scattering data
\begin{equation}
	a(\lambda) = \left(\frac{\lambda-i}{\lambda+i}\right)^2.
\end{equation}
In $(x,t)$ space this corresponds to the following solution
\begin{equation}
	\varphi(x,t) = 4\arctan \frac{t}{\cosh(x-x_0)}
\end{equation}
If $\Pi d <4$ we do not have any discrete degrees of freedom.
\begin{figure*}
    \includegraphics[scale=0.8]{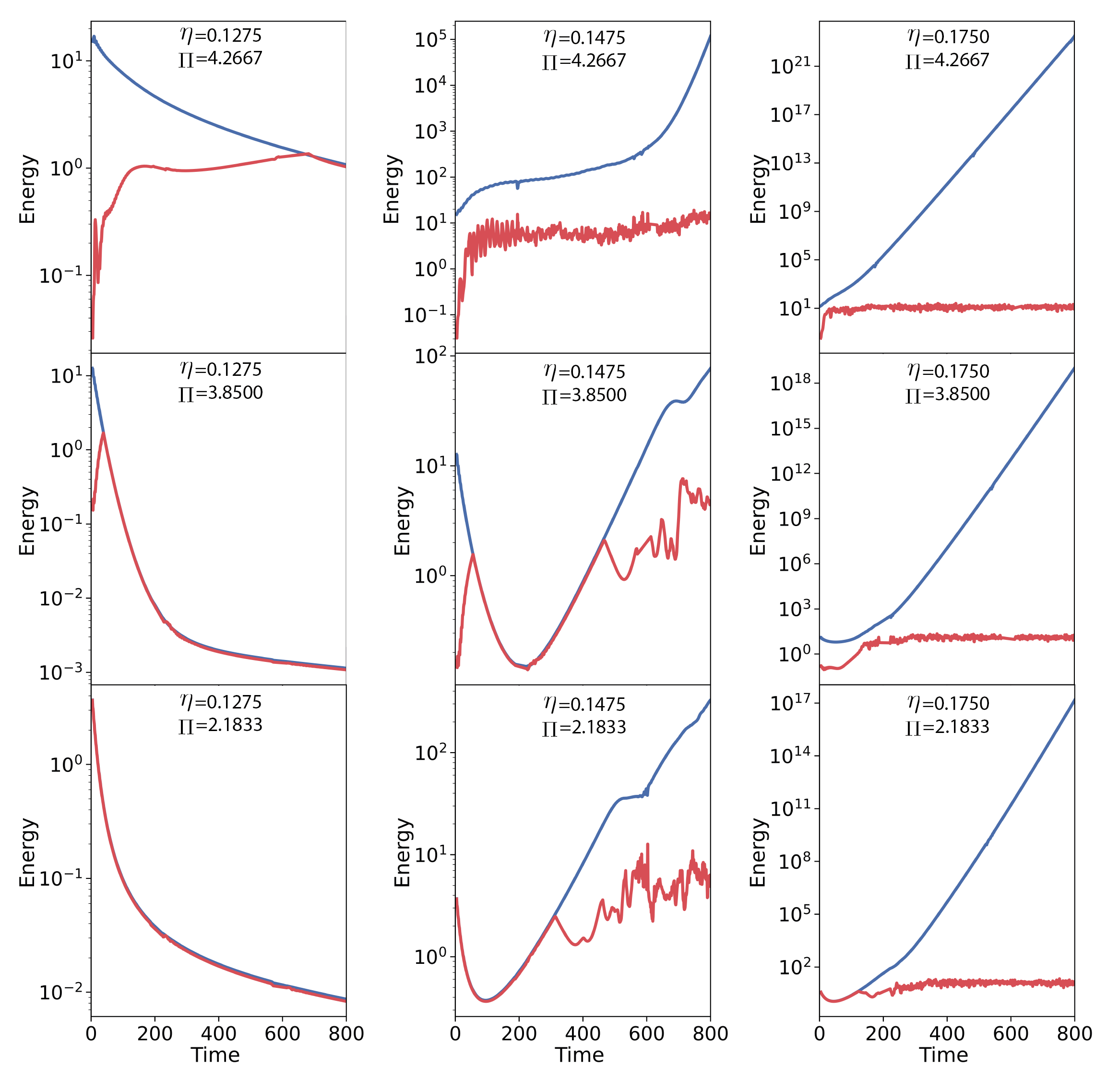}
  \caption{
The full energy $\mathcal{E}_t$ (see Eq. \eqref{etot}) (blue lines)  vs. the radiation energy $\mathcal{E}_r$, given by Eq.
\eqref{erad} (red lines) for the numerical solution of Eq. \eqref{sineG} with the initial condition \eqref{sGprofile} with $
\gamma=0.1$, and the parameters $\eta$ and $\Pi$ shown in each plot.}%  \cc{THE FULL ENERGY OF WHAT EQUATION?}. All simulations are done for $\gamma=0.1$ and each plot is characterized by the pair $(\eta,\Pi)$ where $\eta = {0.1275, 0.1475, 0.175}$ and $\Pi= {2.183, 3.85, 4.27}$. }
  \label{Fig:radiation}
\end{figure*}

\subsection{Conserved quantities}
\label{cq}

Since integrable evolution does not change $a(\lambda)$ its expansion coefficients around some point $\lambda=\lambda_0$
correspond to conserved quantities.
It turns out that if we consider expansions around $\lambda=0$ and $\lambda=\infty$ i.e.
\begin{equation}
    \log a(\lambda) = i \sum\limits_{n=0}^\infty I_{-n}\lambda^n,\quad \log a(\lambda) = i \sum\limits_{n=1}^\infty\frac{I_n}{\lambda^n}
\end{equation}
the coefficients $I_n$ are given by the integrals of local densities.
In particular, $I_0$ is related to the total charge \eqref{q}, and $I_{\pm1}\equiv I_{\pm}$
 are given by
 \begin{equation}\label{Ipm}
	I_{\pm} = \frac{1}{4}\int dx \left(\frac{(\partial_t\varphi\pm\partial_x\varphi)^2}{2} + 1-\cos \varphi\right)
\end{equation}
 Notice that the energy is presented as a sum of these expressions
  \begin{equation}\label{etot}
	\mathcal{E}_t =I_+ + I_-
\end{equation}
Using presentation \eqref{apres} we obtain
 \begin{equation}
	I_\pm = I_\pm^{\rm rad} + I_\pm^{\rm d},
\end{equation}
where the radiation part is given by
 \begin{equation}
 	I_{-}^{\rm rad} =\int\limits_0^\infty\log\frac{1}{|a(\lambda)|^{2}} \frac{d\lambda}{\pi} ,\qquad
 		I_{+}^{\rm rad} = \int\limits_0^\infty\frac{1}{\lambda^2}\log\frac{1}{|a(\lambda)|^{2}} \frac{d\lambda}{\pi}.
 \end{equation}
 The energy of radiation can be presented as
 \begin{equation}\label{erad}
   \mathcal{E}_r \equiv  I_{-}^{\rm rad}+	I_{+}^{\rm rad}   =\int\limits_0^\infty\log\frac{1}{|a(\lambda)|^{2}} \frac{1}{\pi}d \left(\lambda - \frac{1}{\lambda}\right)
 \end{equation}
The discrete part can be expressed as
\begin{equation}
	I^{\rm d}_- = 2\sum {\rm Im} \lambda_i,\qquad I^{\rm d}_+ = -2\sum {\rm Im} \lambda^{-1}_i
\end{equation}
Here by $\lambda_i$ we understand every sort of zero in the upper half plane. This way, every kink is counted as $\lambda_i= i \kappa_i$, while every breather
comes with $\lambda_i$ and $-\bar{\lambda}_i$.
By computing $a(\lambda)$ numerically we can quantify the fraction of radiation in each profile.
To do this we plot in Fig. \eqref{Fig:radiation} the full energy $\mathcal{E}_t $ and the radiation energy $\mathcal{E}_r$.
We clearly see three possible scenarios: (i) in the decaying regime (the left column) where discrete degrees of freedom vanish and the profile is determined completely by radiation $\mathcal{E}_r \approx \mathcal{E}_t$;
(ii) in the exploding regime (the right column) we see that the discrete degrees of freedom take the most part of the profile's energy $\mathcal{E}_r \ll \mathcal{E}_t$; (iii) finally in the central column we see
examples of how discrete degrees of freedom are created from the radiation i.e. we have the regime (i) followed by a revival, followed by the regime (ii) in the course of evolution.
This way, we see that the decay of the solution is possible only if the full profile turns itself completely into the radiation modes, while
the amplification is possible only if the discrete modes are present or are created from the radiation.

Let us also mention the conserved quantities for the Ablowitz-Ladik discretization \eqref{AL}. 
The discrete analogs of the ``number of particle'' and the ``momentum'' reads as 
\begin{equation}
    Q_0[\psi] = \frac{1}{a^2}\sum\limits_{n=-\infty}^\infty \ln (1+ a^2|\psi_n(t)|^2),
\end{equation}
\begin{equation}
    Q_1[\psi] = \sum\limits_{n=-\infty}^\infty i (\psi^*_{n+1}\psi_{n}-\psi^*_{n}\psi_{n+1}).  
\end{equation}
Evaluated on the one-soliton solution \eqref{1solD}, this  quantities give
\begin{equation}
 Q_0[\psi^{(s)}] = \frac{2\mu}{a^2},\qquad 
  Q_1[\psi^{(s)}] = \frac{4\sin(k)\sinh(\mu)}{a^2}. 
\end{equation}
Computing the evolution of $Q_0$ and $Q_1$ with respect to the modified equation \eqref{AL}, we obtain 
 equations \eqref{uD}, \eqref{vD}, with 
 \begin{multline}\label{uD1}
X^{(u)}_a(u,v) = \frac{\tanh(au)}{a^3 u} \eta \sin(av) \left(\sinh(au) + \frac{a u }{\cosh(au)}\right) \\ - \frac{2\Gamma \tanh(au)}{a^3}(1+a^2)  + \frac{2\Gamma}{a^3} \sinh(au) \cos(av)  +  V^{\rm PN}_u 
\end{multline}
\begin{multline}\label{vD1}
    X^{(v)}_a(u,v)  = \frac{\sinh (a u) \eta   \cos (a v) }{a^3 u}\left(1-\frac{2a u}{\sinh(2au)} \right) \\ +2\frac{\sinh (a u)  \sin (a v)\Gamma   (1-a u \coth (a u)) }{a^4 u} +  V^{\rm PN}_v. 
\end{multline}
Here the Pierls-Nabarro potentials are given by 
\begin{equation}
    V^{\rm PN}_u = \frac{2 \pi ^2 \eta  \sinh (a u) \tanh (a u)\sin (a v) \cos \left(\frac{2 \pi  x_c}{a}\right)}{a^4 u^2 \sinh\left(\frac{\pi ^2}{a u}\right) },
\end{equation}
\begin{equation}
     V^{\rm PN}_v = \frac{2 \pi ^2 \sinh (a u) \cos \left(\frac{2 \pi  x_c}{a}\right) (2 \Gamma  \sin (a v)+a \eta  \cos (a v))}{a^5 u^2\sinh\left(\frac{\pi ^2}{a u}\right) }.
\end{equation}
Numerically, we convince ourselves that these terms are not relevant for the dynamics.

% For the perturbed system, we can compare the full energy time dependence with the discrete data contribution extracted directly from the spectral analysis of the profile, in the simplest
% topologically trivial case i.e.
% \begin{equation}\label{d2}
%     \mathcal{E} = 2 {\rm Im} (\lambda_1+\lambda_2 -1/\lambda_1-1/\lambda_2).
% \end{equation}
% The comparison is shown in \eqref{SI-Fig2}, which indicates that the role of radiation is negligible on the background given by the discrete degrees of freedom.

%\begin{figure}[t!]
%\includegraphics[width = 1.0 \columnwidth]{FIg_SI_7.png}
%\caption{Energy dependence of the soliton profile. Solid lines represent numeric integration of the profile \eqref{Ipm}, and dashed lines are contributions from the discrete modes only \eqref{d2}.  }
%\label{SI-Fig2}
%\end{figure}

\subsection{Nonlinear Schr\"odinger equation}
\label{nseP}

For completeness let us also briefly review inverse scattering transformation (IST) for the Nonlinear Schr\"odinger equation and outline the main steps in the derivation of equations \eqref{ffu}, \eqref{ffv} and \eqref{ff1}.
Let us denote the right hand side of Eq. \eqref{nls0} as $P[\psi]$.  For $P[\psi]=0$ the dynamics can be rewritten as the zero-curvature condition
\eqref{zeroC}, with matrices $U$ and $V$, now given by
\begin{equation}\label{alp3}
U  =\left(\begin{array}{cc}
\frac{\lambda}{2i} & i\bar{\psi} \\
i\psi &-\frac{\lambda}{2i}
\end{array}\right),\quad
V = \left(\begin{array}{cc}
-i|\psi|^2 & \partial_x\bar{\psi} \\
-\partial_x\psi & i|\psi|^2
\end{array}\right) - \lambda U.
\end{equation}
Similar, to the IST scheme for sine-Gordon we define the Jost solutions $T_{\pm}$ and the transfer matrix $T(\lambda)$ that connects them.
It can be written as  ({\it cf.} \eqref{ttSG})
\begin{equation}\label{tt}
T(\lambda) = \left(
\begin{array}{cc}
a(\lambda) & - \bar{b}(\lambda) \\
b(\lambda) & \bar{a}(\lambda)
\end{array}
\right).
\end{equation}
The evolution given by Eq. \eqref{nls0} for $\eta=\gamma=0$ is extremely simple on the scattering data \cite{Faddeev_1987}
\begin{equation}\label{dyn1}
a(\lambda,t) = a(\lambda,0),\qquad b(\lambda,t)= b(\lambda,0) e^{-i\lambda^2 t}.
\end{equation}
This follows from the zero curvature conditions \eqref{zeroC} and is also described below.
The solitons correspond to the zeroes of $a(\lambda)$ in the upper half plane, $a(\lambda_i)=0$, ${\rm Im} \lambda_i>0$. Contrary to the sine-Gordon there
is only one type of soliton solutions.
The associated scattering data is the proportionality coefficient between the first column of $T_-$ and the second column of $T_+$ as in Eq. \eqref{gammaSG}.
The evolution of this coefficient is given by
\begin{equation}\label{dyn2}
\mathfrak{g}_j(t) = \mathfrak{g}_j(0)e^{-i\lambda^2_j t}
\end{equation}
The general form of the transmission coefficient $a(\lambda)$ is given by
\begin{equation}
	a(\lambda )  = \prod\limits_{i=1}^k \frac{\lambda- \lambda_i}{\lambda - \bar\lambda_i}  \exp\left(
	\frac{1}{2\pi i} \int \frac{\log (1- |b(\mu)|^2)}{\mu - \lambda-i0}d\mu
	\right)
\end{equation}
The one-soliton solution corresponds to
\begin{equation}
a(\lambda) = \frac{\lambda- \lambda_0}{\lambda-\lambda_0^*},\qquad b(\lambda) = 0.
\end{equation}
for $\lambda_0= v+ iu$, $u>0$, the corresponding profile for $\mathfrak{g} = e^{x_c u + i\eta}$ is given by Eq. \eqref{profS}.

Now let us turn our attention to the case when the non-reciprocity and dissipation are present in the system i.e. $P[\psi]\neq 0$.
In this case the zero curvature condition does not hold anymore; it is transformed into
\begin{equation}
\partial_t U - \partial_x V +[U,V]=\left(\begin{array}{cc}
0 & - P^*[\psi] \\
P[\psi] & 0
\end{array}\right)\equiv \mathcal{P}[\psi]
\end{equation}
One can show using standard ideas of the soliton perturbation theory \cite{Kaup,Karpman2,RevModPhys.61.763} that such a change of the zero curvature condition has the following effect on dynamics of the continuous and discrete scattering data
 \begin{equation}\label{pert1}
\partial_t T(\lambda) - \frac{i\lambda^2}{2} [\sigma_z ,T(\lambda)] =\int\limits_{-\infty}^{\infty} dz T_+^{-1}(z)\mathcal{P}[\psi(z)]T_-(z)
\end{equation}
\begin{multline}
\label{pert2}
\frac{d\mathfrak{g}}{dt} + i \lambda_0^2 \mathfrak{g}  = \\ \frac{i}{\dot{a}(\lambda_0)}\int\limits_{-\infty}^{\infty} dz   \left[\dot{T}^{(1)}_-(z) -\mathfrak{g}\dot{T}^{(2)}_+(z) \right]^{T}\sigma_2\mathcal{P}[\psi(z)]T^{(1)}_-(z)
\end{multline}
\begin{equation}\label{pert3}
\frac{d\lambda_0}{dt}  = \frac{i}{\dot{a}(\lambda_0)}\int\limits_{-\infty}^{\infty} dz  \left[T_+^{(2)}(z)\right]^{T}\sigma_2\mathcal{P}[\psi(z)]T^{(1)}_-(z)
\end{equation}
here the dot means the derivative with respect to the spectral parameter $\lambda$ and the right hand side of Eqs. \eqref{pert2}
and \eqref{pert3} should be evaluated at $\lambda =\lambda_0$.
Notice that the evolution \eqref{dyn1} and \eqref{dyn2} readily follows if one neglects the right hand side in Eqs. \eqref{pert1}-\eqref{pert3}.
To get Eqs. (\ref{ffu}-\ref{ff1}) we use the Jost functions and profile that correspond to one-solitonic profile \eqref{profS}.

\begin{figure}
\includegraphics[width = 1.0 \columnwidth]{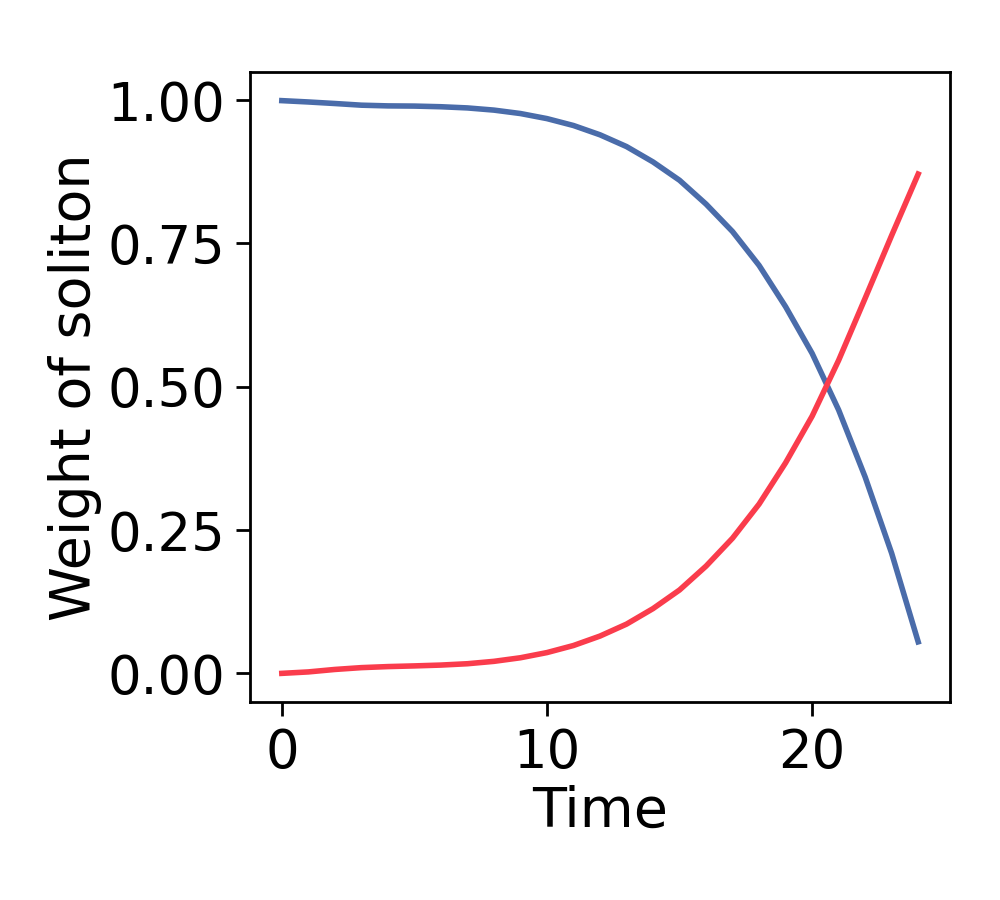}
\caption{The relative weight of a soliton $u/I_0$ (blue line) vs the radiation weight $I_r/I_0$ (red line). Simulation is done for $\eta =0.133$ and $\Gamma=0.1$, with initial conditions $u=1.0$ and $v=0.0$. }
\label{SI-nlsRad}
\end{figure}

The conserved quantities can be deduced from $a(\lambda)$ as coefficients at $\lambda\to \infty$.
The first coefficient in this expansion can be related to the ``total amount of matter'' $I_0= \frac{1}{2}\int_{-\infty}^\infty dx |\psi(x)|^{2}$.
In the case when there is only one soliton (at $\lambda_0 = v+iu$) and radiation the value of $I_0$ reads
\begin{equation}
    I_0 = u + \frac{1}{2\pi} \int\limits_{-\infty}^{\infty} d\lambda \ln\frac{1}{|a(\lambda)|} \equiv u + I_r.
\end{equation}
As we can independently compute $I_0$, $u$, and $I_r$ we can use the above relation to quantify the role of radiation in profile.
We do this in Fig. \eqref{SI-nlsRad} for the perturbed equation \eqref{nls0} for the initial condition $u=1.0$ and $v=0.0$.
In this case, we see that after some time the soliton disappears completely and the whole profile starts to be dominated by the radiation.
This violates presumptions that we used while deriving (\ref{ffu}-\ref{ff1}), which explains deviations of blue lines in Fig. \eqref{Fig4}.

\subsection{Ablowitz-Ladik}

\subsection{Nonlinear Schr\"odinger equation with constant driving}
\label{NLSconstant}
Finally, let us compare the effects of non-reciprocity with the effects of constant driving.
Namely, instead of \eqref{nls0} we consider the following equation \begin{equation}\label{nlsDR}
	i \partial_t \psi +\psi_{xx} + 2|\psi|^2\psi =- f - i \Gamma \psi + \frac{\Gamma }{2}\partial_t \psi .
\end{equation}
Then instead of Eqs. (\ref{ffu}-\ref{ff1}) the evolution of the soliton's parameters read as
\begin{equation}\label{92}
	\frac{dv}{dt} = f\frac{\pi v}{u} \frac{\cos\Phi}{\cosh\frac{\pi v}{2u}}  - \frac{2u^2}{3} \Gamma v,
\end{equation}
\begin{equation}
	\frac{du}{dt} =- \pi f \frac{\cos\Phi}{\cosh\frac{\pi v}{2u}} + u\Gamma(u^2-v^2-2),
\end{equation}
\begin{equation}
	\frac{d \Phi}{dt} = u^2 + v^2 + \frac{\pi^2 f v\sin \Phi  \sinh\frac{\pi v}{2u}}{u^2\left(\cosh\frac{\pi v}{2u}\right)^2},
\end{equation}
\begin{equation}\label{95}
	\frac{dx_c}{dt} = 2v +\frac{\pi^2 f \sin \Phi  \sinh\frac{\pi v}{2u}}{2u^2\left(\cosh\frac{\pi v}{2u}\right)^2}.
\end{equation}
The critical point of this system can be achieved only for the extremely large value of the force $f\gg1$ and very specific initial conditions.
Indeed, the critical value of the amplitude $u_c$ and $v_c$ are related as $u_c = \sqrt{3(2+v_c^2)}$, and the critical value for the phase $\Phi_c$ can be found from \eqref{92} provided that the force is related to $v_c$ as
\begin{widetext}
\begin{equation}
    f^2 =\frac{4 u_c^4 }{\pi ^4 v_c^2}\cosh ^2\left(\frac{\pi  v_c}{2 u_c}\right) \left(v_c^2 \left(\left(\frac{\pi ^2 \Gamma ^2}{9}+\frac{4}{3}\right) u_c^2+4\right)+\left(2 v_c^2+3\right){}^2/\sinh ^{2}\left(\frac{\pi  v_c}{2 u_c}\right)+9\right)
\end{equation}
\end{widetext}
This way as $v_c\to 0$, $|f| =O(1/v_c^2)$ and, as $v_c \to \infty$ , $|f| = O(v_c^3)$, the minimal value is achieved for $v_c^2 \sim 1$, but even in this case the obtained force is quite large.

This is in contrast to the driving introduced by the Lugiato–Lefever equation, which would correspond to the Nonlinear Schr\"odinger perturbation $(\gamma_R - i \gamma_I) \psi + i f$ instead of the right hand side of Eq. \eqref{nls0} (see~\cite{Lucas_NatComm2017,Yu_NatComm2017,Pernet_NatPhys2022}).  One can easily derive a  dynamical system similar to Eqs. \eqref{92}-\eqref{95}. This system has only a static point
\begin{equation}
= \frac{2\sqrt{\gamma_R}\gamma_I }{\pi f}.
\end{equation}
Provided that  $4\gamma_R\gamma^2_I <\pi^2 f^2$ and $\gamma_R>0$ (see in particular \cite{Herr2015}), which corresponds to a static soliton.

\end{document}